\documentclass[11pt,a4paper]{article}

\usepackage[a4paper,margin=1in]{geometry} 
\usepackage[T1]{fontenc}
\usepackage[utf8]{inputenc}
\usepackage{newpxtext} 
\usepackage[final]{microtype}
\usepackage{setspace}
\usepackage{amsmath,amsthm,mathtools} 
\usepackage{bm}
\usepackage{bbm}
\numberwithin{equation}{section}
\allowdisplaybreaks

\usepackage{newpxmath} 

\usepackage{graphicx}
\usepackage{subcaption}
\usepackage{booktabs}
\usepackage{threeparttable}
\usepackage{adjustbox}
\usepackage{makecell}
\usepackage{siunitx}
\usepackage{enumitem}
\setlist{nosep}

\usepackage[authoryear]{natbib}

\usepackage[dvipsnames]{xcolor}
\usepackage[
  colorlinks=true,
  linkcolor=black, 
  citecolor=MidnightBlue, 
  urlcolor=RoyalBlue 
]{hyperref}
\DeclareRobustCommand{\email}[1]{\href{mailto:#1}{\texttt{#1}}}

\theoremstyle{plain}

\theoremstyle{definition}

\theoremstyle{remark}

\title{Probabilistic Estimation of Hidden Migrant Fatalities Along the Central Mediterranean Route\vspace{0em}}
\author{%
  Gregor Zens\thanks{International Institute for Applied Systems Analysis (IIASA). Corresponding author: \email{zens@iiasa.ac.at}.}%
  \and
  Zoe Sigman\thanks{DYNAMICS, The Hertie School and Humboldt University of Berlin. \email{Z.Sigman@phd.hertie-school.org}.}%
}
\date{September 2026} 

\newcommand{\keywords}[1]{\par\smallskip\noindent\textbf{Keywords:} #1\par}

\newenvironment{acks}[1][Acknowledgments]{\subsection*{#1}}{}
\newenvironment{funding}{\subsection*{Funding}}{}

\begin{document}
\maketitle

\begin{abstract}
\noindent Estimating the number of migrants who die or go missing along dangerous routes such as the Central Mediterranean remains challenging as available records are incomplete. Some incidents are never documented, and fatalities associated with such unobserved incidents are absent from observed totals. We propose a Bayesian approach for probabilistic estimation of total migrant fatalities in such settings. Building on recent developments in multiple-systems estimation, we develop a time-stratified latent-class framework that accommodates missing fatality counts for unobserved incidents. We apply the method to recoded incident-level data from the \textit{Missing Migrants Project} for the Central Mediterranean route from 2014 to 2025, encompassing 25,712 fatalities across 1,562 incidents. Our model yields 95\% credible intervals of 31,922--41,413 fatalities and 2,170--2,952 deadly incidents, indicating that approximately 62\%--81\% of fatalities and 53\%--72\% of incidents are reflected in the available data. We estimate that unreported fatalities were concentrated between 2014 and 2016. Furthermore, we document that reporting likelihood increases with incident severity, implying that smaller incidents are most likely to remain undetected. While contingent on modeling assumptions and incomplete data, our method provides a broadly applicable and principled alternative to naive data adjustment methods.
\end{abstract}

\vspace{0.5em}
\keywords{migrant mortality; multiple-systems estimation; Bayesian inference; mixture models; measurement error}

\section{Introduction}\label{sec:intro}

Deaths along high-risk migration routes such as the Central Mediterranean are widely acknowledged, yet systematically undercounted. Deadly incidents typically occur offshore and outside of routine observation, reporting is fragmented across institutions with different incentives and capacities, and some incidents become known only through partial information from survivors or families. As a result, observed incident records necessarily represent an incomplete subset of all fatal incidents, and recorded fatalities constitute a lower bound on true mortality.

This gap in migrant mortality data carries significant implications across historical, social, and political domains. Maintaining an accurate historical record is fundamental to quantifying the mortality risk and lethality inherent in migratory routes. Beyond historical accuracy, these metrics serve as empirical inputs for evidence-based policymaking. More robust estimates allow for more rigorous monitoring and evaluation of policy impacts -- including search-and-rescue operations -- and enable comparative analyses of how mortality risks fluctuate across different geographic routes, time periods, and border policy regimes. Furthermore, improved methodology for documenting migrant fatalities directly supports the monitoring of the \textit{United Nations (UN) Sustainable Development Goal Target 10.7}, which advocates for safe, orderly, and responsible migration through well-managed policies.

From a statistical perspective, the central challenge in estimating fatalities at sea is that \textit{both} the number of fatal incidents and the fatality counts associated with missed incidents are unobserved. However, most methods in the capture--recapture literature focus primarily on estimating the number of missing \emph{incidents} (or individuals) and do not directly target additional unobserved \textit{marks} of the missing incidents, in our case the mortality burden attributable to incidents that were never recorded.

We propose a Bayesian approach to estimate total migrant fatalities in such settings, which explicitly accounts for unobserved incidents and the fatality counts associated with them. The key methodological idea is to augment latent class models for multiple systems analysis with an additional mechanism to account for unobserved fatalities.
Our model allows for complex dependencies between fatality counts and capture probabilities, as the number of fatalities associated with an incident may impact the likelihood of reporting. The model further allows for flexible dependence structures among the reporting sources. Finally, the proposed mixture formulation explicitly accounts for strata in the data, accommodating the fact that the underlying reporting and fatality patterns may vary across time periods.

We apply the model to manually recoded incident-level data obtained from the \textit{Missing Migrants Project} (MMP) database maintained by the \textit{International Organization for Migration} (IOM) and investigate migrant mortality along the Central Mediterranean route from February 2014 to December 2025. Our main findings can be summarized as follows. First, the baseline model specification yields a 95\% credible interval of 31,922 to 41,413 fatalities (about 7--10 per day) and 2,170 to 2,952 deadly incidents. Compared to the 25,712 fatalities (about 6 per day) and 1,562 incidents recorded in the MMP data, this indicates that the available records capture a majority of the mortality burden -- approximately 62\% to 81\% of total fatalities and 53\% to 72\% of all fatal incidents. However, a substantial share of incidents and fatalities is estimated to go undocumented. Second, our results indicate that the absolute number of unobserved fatalities was likely concentrated between 2014 and 2016. 
Third, we find that the average reporting likelihood increases with incident severity, implying that smaller incidents are the most likely to go undocumented.

\subsection{Related Literature and Contributions}

The proposed methodology builds on a long tradition of capture--recapture and multiple-systems analysis techniques, particularly with respect to frameworks developed within human rights and public policy contexts \citep{lum2013applications, bird2018multiple, manrique2022capture}. These techniques have been successfully applied to estimate the number of casualties in armed conflicts \citep{ball2002killings, lum2010measuring, manrique2013multiple, manrique2019estimating, dahab2025war} and to quantify the prevalence of modern slavery \citep{silverman2020multiple}. Estimates based on such methods have also served as supporting evidence in war crime tribunals \citep{ball2002statistics}. Historically, these approaches have their roots in ecology, where they are widely used to estimate, e.g., animal population sizes; for a comprehensive overview, see \citet{amstrup2010handbook}.

Existing approaches for estimating the size of hidden populations vary in detail and complexity. However, they all fundamentally rely on analyzing inclusion patterns across observations to estimate the probability of non-observation, and by extension the number of unobserved individuals or incidents. The majority of human rights applications utilize frameworks developed for multiple overlapping lists, including log-linear models \citep[e.g.,][]{ball2002killings, silverman2020multiple}, potentially combined with model averaging ideas \citep[e.g.,][]{madigan1997bayesian}. An overview is provided in \citet{manrique2013multiple}. An approach closely related to the present work is \citet{manrique2016bayesian} who proposes to integrate latent class modeling with Dirichlet process mixtures. More recent methodological work in this realm has focused on sparse capture patterns and the resulting inferential challenges \citep{johndrow2019low, chan2021multiple}, and bootstrapping for uncertainty quantification \citep{silverman2024bootstrapping}.

Most existing methods focus on estimation at the level of individuals or single occurrences -- the typical unit of observation in this field -- with less emphasis on joint modeling of incidents and any associated \textit{marks} (in our case, fatalities). However, marked incidents represent the most natural unit of observation in available data sources on migrant fatalities at sea, where deaths are often clustered around vessel failure. To our knowledge, only \citet{farcomeni2022many} addresses this in a related context, estimating migrant fatalities on European borders using a regression-based approach rooted in single-source capture--recapture methodology.\footnote{Magazinnik, Wucherpfennig, and Rodríguez-Sánchez have, in parallel and independently, developed an approach to the joint modeling of incidents and associated marks along the Central Mediterranean Route (presented at the \textit{Missing Migrants Project Estimations Workshop} held at the IOM in Berlin in 2025, with a manuscript in preparation).} While this represents a significant step toward principled statistical inference in such settings, it relies on assumptions that are restrictive, and particularly so within the Central Mediterranean context. First, the method treats reporting sources as independent, even though lists are often operationally linked. For instance, reports of international organizations may rely on official records, while media reports may echo other outlets. Second, the approach does not directly facilitate information sharing across strata, such as years, potentially destabilizing inference in the presence of substantial variations in surveillance intensity and operational capacity across different periods. Third, inference is conditional on the choice of a single parametric specification within a generalized linear model class, in contrast to the flexible semi-parametric fit provided by mixture-based approaches. These limitations motivate the development of methods capable of accommodating source dependence while providing greater flexibility and robustness for stratified settings.

Our contribution to this literature is twofold. First, from a statistical perspective, we develop a Bayesian framework for multiple systems analysis for marked incidents, capable of estimating posterior distributions for both the total number of incidents and any functionals of interest regarding the unreported marks. The framework combines a sparse finite mixture of capture patterns with shifted negative binomial mark kernels, and we derive a posterior simulation algorithm based on data augmentation. Second, in the context of studying migrant mortality, we conduct an in-depth analysis of the Central Mediterranean route, leveraging recoded incident-level data from the MMP covering the period 2014--2025. We produce probabilistic estimates of total mortality and the implied undercount relative to recorded totals. In supplementary analyses, we combine these estimates with additional assumptions and auxiliary data to explore plausible ranges for mortality risk per crossing attempt in this context.

The remainder of the paper is structured as follows. Section~\ref{sec:data} describes the data, scientific questions, and key challenges when working with observed incident records in this context. Section~\ref{sec:methods} presents the model, prior specification, and posterior simulation approach. Section~\ref{sec:results} presents our analysis of migrant deaths en route in the Central Mediterranean from 2014 to 2025. Section~\ref{sec:conclusion} concludes with a discussion of limitations and directions for future research. Additional simulation experiments and further details are provided in the supplementary material.

\section{Data and Scientific Questions}
\label{sec:data}

Our analysis utilizes incident-level records from the MMP database managed by the IOM, which tracks deaths and disappearances of transiting migrants (\citealp{iom_mmp}). The MMP is widely acknowledged as the most comprehensive global inventory of migrant fatalities and serves as a common reference point for related monitoring efforts; for instance, the \textit{UN High Commissioner for Refugees} (UNHCR) applies similar methodology and exchanges information with the IOM to verify figures.\footnote{See IOM's own documentation (\url{https://missingmigrants.iom.int/methodology}) or M. Kessous, ``Immigration to Europe: The `impossible' task of counting deaths at sea'', \textit{Le Monde}, February 23, 2026.} It covers incidents involving at least one fatality. We extracted all such records regarding the Central Mediterranean route available as of March 12, 2026, and restrict our sample to the time period from January 1, 2014 to December 31, 2025. This yields a set of 1,562 temporally localized incidents with well-defined incident-level characteristics that occurred between February 17, 2014 and December 30, 2025.


For each incident, the outcome of interest is the reported total number of people \emph{dead or missing}. We refer to this count throughout as \textit{fatalities}. We define \textit{missing} and \textit{dead} according to the definitions utilized by the MMP.\footnote{See \url{https://missingmigrants.iom.int/methodology} (accessed Feb. 17, 2026).}  In the context of the Central Mediterranean Route, \textit{missing} refers to the number of migrants who are reported by eyewitnesses to have been onboard a shipwrecked vessel and are assumed to have died, but whose remains have not been recovered. \textit{Dead} refers to those migrants who have died and whose remains have been recovered. After data processing, the total fatality count in the data is 25,712.\footnote{The MMP updates its data regularly. Hence, future incident and fatality counts may vary.}

In accordance with the MMP's inclusion criteria, the data analyzed here exclude deaths that occur when migrants are not in transit, e.g., in formal or informal camps or detention centers, work-related deaths of migrant workers, and deaths of migrants who have reached their destination country or have settled in a destination for more than two weeks. They also exclude reports of those who have gone missing during migration, but whose deaths cannot be assumed, e.g., those people who have lost contact with loved ones for unverifiable reasons, such as detention. Our analysis should be interpreted accordingly as referring only to the population of migrants who have died during transit.

\subsection{List Creation}
\label{sec:lists}

For each incident in the database, the MMP team collects a set of \textit{information sources} which have reported on a given incident. We classify each unique information source string in the MMP data into one of four categories representing four surveillance systems with relatively distinct coverage, dynamics, and capabilities. 
The first category encompasses sources originating directly from the UN system and other intergovernmental or international organizations, including reports from the IOM itself, the \textit{UN High Commissioner for Refugees}, and related bodies. The second category consists of official governmental sources, such as coast guard agencies, national ministries, and other state institutions. The third category includes non-governmental organizations and civil society actors, ranging from search-and-rescue operations to advocacy groups and humanitarian organizations. The fourth category comprises media sources, including both traditional news outlets and aggregation services.

We adopt these relatively broad groupings to balance two competing considerations. On the one hand, finer distinctions -- for instance, separating local from national or international media -- would yield more lists and more homogeneous reporter categories, but risk producing sparse and highly correlated lists. On the other hand, excessively coarse categories would obscure meaningful differences in reporting behavior. Our four-category scheme aims to identify source types that function as relatively independent observation and surveillance systems. Because any such coding necessarily involves judgment, we additionally construct a coarser three-list variant that merges the UN/IGO and NGO/Humanitarian categories, and a finer five-list variant that splits official sources into European and non-European institutions. These alternative aggregations are used throughout our empirical analysis as robustness and sensitivity checks on the list construction itself.


\begin{table}[t]
\footnotesize
\centering
\begin{tabular}{p{0.5\textwidth}cccc}
\toprule
Raw source string (MMP) & UN/IGO & Official & NGO/Hum. & Media \\
\midrule
IOM Libya, Lana News & 1 & 0 & 0 & 1 \\
Ansa, UNHCR Lampedusa & 1 & 0 & 0 & 1 \\
Sea Punks & 0 & 0 & 1 & 0 \\
IOM Libya, Pakistan PM Shehbaz Sharif & 1 & 1 & 0 & 0 \\
RESQSHIP, EFE & 0 & 0 & 1 & 1 \\
\bottomrule
\end{tabular}
\caption{Examples of information source classification and derived capture indicators. Each raw source string in the MMP data is coded into binary inclusion indicators for four derived reporting systems (UN/IGO, Official, NGO/Humanitarian, Media).}
\label{tab:coding_examples}
\end{table}

Tab.~\ref{tab:coding_examples} presents examples of our coding logic, illustrating how specific source strings map to the four categories. Although tedious, coding these source strings is mostly straightforward. However, certain cases unavoidably require judgment calls. For instance, \textit{ReliefWeb} is operated under the \textit{UN Office for the Coordination of Humanitarian Affairs} but functions primarily as an aggregator of content from diverse origins, including both UN agencies and independent news organizations. We classify such online aggregators as media sources, prioritizing their functional role over their institutional affiliation. Although no classification scheme can be entirely unambiguous, we have aimed for a procedure that is conservative, internally consistent, and reproducible.\footnote{Zoe Sigman previously served on the MMP data collection team for the Central Mediterranean, which informed various aspects of the coding and modeling decisions described in this paper.} 

We emphasize that this recoding is \textit{not} intended or claimed to render the resulting lists independent. In fact, dependence between the derived lists is expected -- for example, media coverage may draw on government press releases, inducing positive dependence between those two capture systems -- and is explicitly accommodated in the modeling framework introduced below. 

\subsection{Benefits and Limitations of the Meta-Observer Approach}
\label{sec:metaobserver}

It is important to acknowledge that the reporter categories used here do not constitute independently operating ``lists'' in the classical capture--recapture sense. Rather, they are defined by recoding information sources as recorded by a \textit{meta-observer} -- here, the MMP team -- which collects and verifies reports from multiple surveillance systems. 

The key advantage of the recoding approach is that it enables us to model correlations between reporters (and, eventually, fatality marks) using ideas from multiple-systems analysis. A central caveat of this approach, however, is that the \textit{meta-observer} may introduce bias in how it aggregates information from the underlying data generators. While it is plausible to assume that the IOM observation process credibly spans all four surveillance systems -- and that incidents recorded in the database were indeed reported by the stated sources -- it may also occur that not all sources are consistently documented. For instance, if an incident enters the database via the UN system, an additional media report might not always be recorded even if it existed. Under this perspective, non-reporting of a list is subject to uncertainty. That is, the absence of a source label does not necessarily imply that the source failed to report the event. By contrast, sources that \textit{are} recorded in the MMP system correspond to reports that were verified and documented by the MMP team prior to inclusion. 

A direct implication of this measurement system concerns the estimand. Since all information passes through the MMP retrieval process, the natural -- and, in our view, the only well-defined -- estimand in this setting is the number of incidents and fatalities that remain \textit{undocumented by the IOM MMP}. All estimates reported below should be interpreted accordingly. They quantify the gap between the MMP record and the (latent) complete record that the MMP system targets.

That said, in the context of migrant mortality en route, a \textit{meta-observer} structure of this kind is nearly always the only feasible -- and typically also the best -- available data source. As a result, alternative approaches face the same limitation (e.g., \citealp{farcomeni2022many}). The relatively broad capture systems we operationalize do, however, lend our approach a certain degree of robustness. For instance, it is more plausible that the MMP team captures \textit{any} media report on a given incident than that they capture \textit{every} such report. In our analyses below, we assume that the MMP retrieval process is approximately comprehensive with respect to the four surveillance systems.
We revisit these issues in Sec.~\ref{sec:conclusion}, where we highlight sources of bias and avenues for future work, including targeted audits of individual incidents and model extensions that explicitly account for uncertainty in source recording.

\subsection{Research Questions}

Drawing on the dataset described above, we address three primary research questions. First, what is the total number of migrant fatalities and deadly incidents on the Central Mediterranean route between 2014 and 2025? Second, how do these figures evolve over time, specifically regarding annual trends and intra-year seasonal patterns? Third, what is the relationship between the severity of an incident and its probability of being captured by surveillance systems? 

Each of these questions requires knowledge about the full event process, including incidents and fatalities that are not observed in any list. Our strategy for answering these questions is to build a fully probabilistic estimation framework that imputes both missing incidents and their fatality counts in the estimation process. To motivate some key modeling choices, the next subsection reports some descriptive characteristics of the data that we will explicitly take into account when introducing the statistical modeling framework in Sec.~\ref{sec:methods}.

\subsection{Empirical Stylized Facts}

We begin by describing basic patterns regarding fatalities and incidents in the reported data (Fig.~\ref{fig:incidents_fatalities_over_time}). Both the mean and the variance of the fatality distribution decline over the study period. In 2014, the average number of fatalities per incident is almost 20, decreasing to less than 5 in 2025. Similarly, total reported fatalities decrease from a peak of 4,572 in 2016 to 1,286 in 2025. Overall, the distribution of fatalities is right-skewed, with 35\% of incidents including only a single fatality. Across the full sample, the mean is 16.5 fatalities per incident, while the maximum corresponds to an event with 1,022 reported fatalities on April 18, 2015. 

Conditional on an incident being observed, the UN/IGO list exhibits the highest average capture rate (72\%), followed by the media (30\%), NGOs and humanitarian partners (17\%), and official sources (11\%).\footnote{Counts of incidents for all observed reporting patterns iare provided in Supplementary Tabs.~\ref{tab:mse_overlap}, \ref{tab:mse_overlap_3L}, and \ref{tab:mse_overlap_5L}.} Reporting patterns vary substantially across years (Fig.~\ref{fig:reporting_by_list}). In 2014--2015, the media list is the strongest source, while the UN/IGO list contributes relatively little. Over time, this relationship reverses. Pairwise reporting dynamics also change substantially -- for example, the UN/IGO--official pair exhibits a positive trend over the years. These patterns partially reflect the methods of the MMP team, which relied primarily on media reports and disembarkation reports from IOM Italy until 2017. After 2017, there was a shift towards more reports from NGOs and other IGOs, the initiation of reports from IOM Libya, 
as well as a more dedicated staffing for the MMP team.


Finally, the data indicate a complex marginal relationship between incident severity and reporting patterns (Fig.~\ref{fig:reporting_by_list}). For instance, the UN/IGO list exhibits a U-shaped pattern. The other lists show similarly nuanced behavior. For example, the media list has comparatively low reporting probabilities for the smallest incidents, while official sources report the largest incidents less frequently. Importantly, because these patterns are aggregated over the entire study period, they may reflect temporal shifts in reporting practices and should thus be interpreted with caution.  

Taken together, these stylized facts imply a minimum set of three requirements for a model capable of addressing the research questions outlined above. First, the model must allow for time-stratified inference for both the mark distribution and reporting patterns.\footnote{See also \url{https://hrdag.org/2013/03/20/mse-stratification-estimation/} (accessed Feb. 9, 2026) on the importance of stratification in multiple systems analysis.} Second, the model should accommodate potentially complex relationships between reporters, as well as between fatalities and reporters, that may also vary over time. Third, the model must account for the integer-valued, right-skewed, and overdispersed distribution of fatalities per incident. In Sec.~\ref{sec:methods}, we develop a probabilistic model that addresses these challenges jointly.

\section{Statistical Methodology}
\label{sec:methods}

We consider a population of incidents partitioned into $G$ mutually exclusive strata (e.g., time periods or geographic regions). Let $N_g$ denote the unknown total number of incidents in stratum $g \in \{1,\dots,G\}$. We observe a subset of these incidents through $R$ reporting systems.\footnote{We treat the number and identity of the reporting systems as constant across strata. In our application this is appropriate by construction, as all source systems underlying the derived lists are active throughout the entire study period, so the same $R$ capture systems operate in every stratum. In settings where lists enter or exit over time, stratum-specific list configurations would be required.} For each observed incident $i$, we record its stratum indicator $h_i \in \{1,\dots,G\}$, a reporting pattern $\mathbf{s}_i=(s_{i1},\dots,s_{iR}) \in \{0,1\}^R$ with $s_{ij}=1$ if system $j$ reported incident $i$ (and $0$ otherwise), and a positive integer-valued mark $y_i\in\mathbb{N}_+$ measuring incident magnitude. In the present application, $y_i$ is the reported fatality count, which is at least one for every incident in the database. The observed dataset contains $m$ incidents and is \emph{zero-truncated} by construction, since incidents that are not reported by any system (i.e., with $\mathbf{s}_i=\mathbf{0}$) are unobserved. The analysis targets, for each stratum $g$, both the total number of incidents and the total mark sum. Writing $m_g$ for the observed number of incidents in stratum $g$ and $n_{0g}$ for the (unknown) number of unobserved incidents in that stratum, we have
\begin{equation}
    N_g \;=\; m_g + n_{0g}.
\end{equation}
The total mark sum -- in our case, the total fatality count -- in stratum $g$ is
\begin{equation}
    Y_g^{\text{tot}}
    \;=\;
    \underbrace{\sum_{i \in \mathcal{O}_g} y_i}_{\text{observed}}
    \;+\;
    \underbrace{\sum_{\ell=1}^{n_{0g}} y_{0g\ell}}_{\text{unobserved}},
\end{equation}
where $\mathcal{O}_g$ denotes the set of observed incidents in stratum $g$, and $y_{0g\ell}$ is the (unknown) mark of the $\ell$-th unobserved incident in stratum $g$. For each stratum $g$, the number of unobserved incidents $n_{0g}$ is treated as a random variable. We assume the total population $N_g$ in stratum $g$ follows a Poisson model $N_g \sim \text{Poisson}(\lambda_g)$ with intensity $\lambda_g$.

\subsection{Stratified Mark-Augmented Latent Class Model}

Motivated by the success of latent class models for multiple systems analysis (\citealp{manrique2016bayesian}), we account for unobserved heterogeneity, as well as list--list and list--mark dependencies, using a Bayesian latent class mixture model. We assume there exist $K$ latent clusters of incidents. Let $z_i \in \{1, \dots, K\}$ be the latent cluster assignment for incident $i$. Within cluster $k$, captures are independent Bernoulli events. The probability that incident $i$ is captured by list $j$ is $p_{kj}$. The probability of observing the capture pattern $\mathbf{s}_i$ is thus
\begin{equation}
    P(\mathbf{s}_i \mid z_i = k, \mathbf{p}_k) = \prod_{j=1}^R p_{kj}^{s_{ij}} (1 - p_{kj})^{1 - s_{ij}},
\end{equation}
where $\mathbf{p}_k = (p_{k1}, \dots, p_{kR})$. Consequently, the probability that an incident in cluster $k$ remains completely unobserved is $q_k = \prod_{j=1}^R (1 - p_{kj})$. 

We assume that within each cluster the fatality counts follow a \textit{shifted negative binomial} (SNB) distribution. Writing $y_i = 1 + x_i$, we assume
\begin{equation}
    x_i \mid z_i = k \sim \mathrm{NB}(r_k, p^{\mathrm{nb}}_k), \qquad 
    \Pr(x_i = x \mid r_k, p^{\mathrm{nb}}_k) = \frac{\Gamma(x + r_k)}{\Gamma(r_k)\, x!}\,(1 - p^{\mathrm{nb}}_k)^{r_k}\,(p^{\mathrm{nb}}_k)^{x},
    \label{eq:snb}
\end{equation}
with cluster-specific size parameter $r_k > 0$ and probability parameter $p^{\mathrm{nb}}_k \in (0,1)$.\footnote{We adopt the parameterization of \citet{zhou2015negative}, under which the conjugate data augmentation scheme used for posterior simulation is most transparent.} We denote the resulting probability mass function of $y_i$ by $f_{\mathrm{SNB}}(y_i \mid r_k, p^{\mathrm{nb}}_k)$. The shift by one reflects that every recorded incident involves at least one fatality, so the model matches the support of the data. The implied cluster-specific mean is $1 + r_k\, \omega_k$ with odds $\omega_k = p^{\mathrm{nb}}_k / (1 - p^{\mathrm{nb}}_k)$, and the variance $r_k\,\omega_k / (1 - p^{\mathrm{nb}}_k)$ exceeds the mean of the count component, so each cluster can be substantially overdispersed.

The parameters of these clusters are shared across all strata to facilitate information sharing and stability of the model. To allow for stratum-specific inference, we assume the mixing proportions of the mixture model vary by stratum. 
Hence, the latent cluster assignments are drawn from a stratum-specific discrete distribution. For an incident in stratum $g$, we have
\begin{equation}
    P(z_i = k \mid h_i = g) = \pi_{gk}, \quad \text{with } \sum_{k=1}^K \pi_{gk} = 1.
\end{equation}

While we assume conditional independence of capture events and marks within each cluster, the marginal distribution induced by the mixture allows for rich and complex dependencies between lists and marks. By averaging over $K$ distinct clusters, the model effectively captures the correlation structures present in the aggregate data \citep{mclachlan2000finite, hagenaars2002applied, fruhwirth2006finite, manrique2016bayesian}.

\subsection{Interpretation and Rationale of the Model Structure}

Beyond providing a flexible approach for density estimation, the latent class formulation reflects the operational realities in the context of the Central Mediterranean route. Here, latent classes naturally represent unobserved heterogeneity arising from complex environmental and logistical factors that simultaneously drive incident lethality and the likelihood of documentation. For instance, incidents occurring deep within the Libyan Search and Rescue zone or during severe weather conditions may exhibit both higher fatality rates due to delayed rescue and lower detection probabilities by European official or media sources. Similarly, the type of vessel involved -- ranging from fragile rubber dinghies to large, densely packed fishing boats -- creates distinct casualty and reporting signatures. While the capsizing of a large vessel may trigger a multi-agency response that activates multiple surveillance systems, sinkings involving smaller vessels may remain largely invisible, becoming known only through survivor testimonies collected by humanitarian agencies. By segmenting incidents into latent clusters, our model explicitly accounts for these dynamics. This allows the framework to capture the dependence between incident severity and reporting patterns without requiring granular covariate data, which is frequently missing or entirely unobservable, even for known incidents.

The decision to hold cluster-specific parameters constant across strata while allowing the mixture weights ($\pi_{gk}$) to vary is driven by two main considerations. First, from a statistical and computational perspective, it is a parsimonious choice that allows the model to learn a shared ``dictionary'' of latent components. By sharing the component parameters across strata, the model borrows strength from the entire dataset. At the same time, the stratum-specific weights allow the model to flexibly turn specific clusters ``on'' or ``off'' in any given stratum, avoiding potential overparameterization due to estimating entirely separate models per stratum. Second, from an application perspective, this structure accommodates a wide range of real-world dynamics. It allows certain incident ``archetypes'' -- for instance low-visibility, localized rubber dinghy deflations -- to be persistently present across all strata. Simultaneously, other archetypes can fluctuate in prevalence, emerge as highly stratum-specific events, or even split into distinct components -- for instance, if reporting systems become more adept at documenting high-fatality incidents in later years. Consequently, the proposed latent class formulation effectively captures both persistent, long-term surveillance patterns and transient, period-specific shifts.

Note that, in principle, one could alternatively ``expand'' the incident-level data by replicating each incident's capture pattern once per reported fatality (e.g., an incident with 5 fatalities yields 5 identical rows) and apply standard capture-recapture methodology. This would effectively treat fatalities as independent observational units with the same list-inclusion pattern, thereby reweighting the likelihood toward high-fatality incidents and implicitly assuming within-incident independence and homogeneous capture across deaths. Because our lists are best interpreted as incident-level reporters (rather than individual-level registries), such expansion constitutes pseudo-replication, can distort estimation of capture probabilities, and may understate uncertainty. We therefore prefer to model fatalities as an incident mark jointly with the capture pattern, which preserves the correct observational unit and allows detection to depend on incident severity.

\subsection{Prior Elicitation}
\label{sec:priors}

We employ a sparse finite mixture approach \citep{malsiner2016model} to flexibly infer both the number of active clusters and their associated weights. The key ingredient is a prior on the mixture weights $\boldsymbol{\pi}_{g}$ that induces \textit{sparsity}, such that only a small subset of the $K$ available components is active \textit{a posteriori}. This allows the effective number of mixture components required to represent dependence patterns in the data to be treated as an unknown quantity and learned within the estimation procedure. Rather than fixing a single number of active clusters, this method directly accounts for the uncertainty across the true number of clusters. Closely related ideas motivate the Dirichlet process mixture specification used by \citet{manrique2016bayesian}. We expect broadly similar results under either mixture construction, since recent work suggests that the distinction between sparse finite and Dirichlet process mixtures is typically less influential than the choice of within-model prior specification \citep{fruhwirth2019here}. The full hierarchical prior structure we use is specified as follows:

\begin{align}
    \boldsymbol{\pi}_g \mid \alpha_g &\sim \text{Dirichlet}\left(\frac{\alpha_g}{K}, \dots, \frac{\alpha_g}{K}\right), \label{eq:dirichlet} \\
    \alpha_g &\sim \text{Gamma}(a_\alpha, b_\alpha), \\
    p_{kj} &\sim \text{Beta}(a_p, b_p), \\
    p^{\mathrm{nb}}_k &\sim \text{Beta}(a_\omega, b_\omega), \\
    r_k &\sim \text{Gamma}(a_r, b_r), \\
    p(\lambda_g) &\propto \lambda_g^{-1}. \label{eq:lambdaprior}
\end{align}

The concentration parameters $\alpha_g$ are estimated from the data, allowing the model to adapt the effective model complexity per stratum. For the cluster-specific mark distributions, we use weakly informative priors that are centered via a simple closed-form rule. Specifically, we choose the hyperparameters of the Beta prior on $p^{\mathrm{nb}}_k$ such that
\begin{equation}
    a_\omega = (b_\omega - 1)\,\frac{2\,(G - 1)}{\mathbb{E}[r_k]}, \qquad
    G = \exp\!\left(\tfrac{1}{m}\textstyle\sum_{i=1}^m \log y_i\right),
    \label{eq:centring}
\end{equation}
where $G$ is the geometric mean of the observed marks. This rule sets
$\mathbb{E}[r_k]\,\mathbb{E}[\omega_k] = 2(G-1)$, with odds $\omega_k = p^{\mathrm{nb}}_k/(1-p^{\mathrm{nb}}_k)$, pinning the prior predictive
mean mark at $\mathbb{E}[y] = 1 + \mathbb{E}[r_k]\,\mathbb{E}[\omega_k] = 2G - 1$
for any choice of the prior on $r_k$. It thereby anchors a typical prior draw around the geometric mean of the observed marks, a location measure that is robust to the small number of mass-casualty events in our data. Note that centering the mark prior on the geometric mean of the \textit{observed} marks renders this component of the prior specification empirical Bayes, as the hyperparameter $a_\omega$ is determined by a summary statistic of the observed data. 

Finally, for the population intensity $\lambda_g$, we employ the improper scale prior in Eq.~(\ref{eq:lambdaprior}), obtained as the limit of the conjugate $\text{Gamma}(a_\lambda, b_\lambda)$ family as $a_\lambda, b_\lambda \to 0$. This is the standard scale-invariant choice, assigning equal prior weight to equal proportional changes in population size, and marginally implies the prior $P(N_g = n) \propto 1/n$, $n \geq 1$, on the latent number of events \citep{wang2007bayesian}. The corresponding full conditional distribution of $\lambda_g$ is proper whenever at least one incident is observed in stratum $g$.

\subsection{Hyperparameter Choices}
\label{sec:hyperparameters}

Throughout, we utilize the following baseline hyperparameter choices. We set the upper bound on the number of clusters to $K=100$, large enough to ensure it does not constrain the posterior distribution of active components heavily. For the capture probabilities, we select $a_p = b_p = 1$, implying a uniform prior on $[0,1]$ that assumes no strong prior knowledge regarding the sensitivity of the reporting lists. To govern the sparsity of the finite mixture, we place a $\text{Gamma}(2, 1)$ prior on the concentration parameters $\alpha_g$. A shape parameter of two keeps prior mass away from $\alpha_g = 0$, so redundant components can still be emptied \textit{a posteriori}, but the mixture is discouraged from collapsing onto a single component. This is the desired behavior for the density estimation task at hand. For the mark distributions, we set $r_k \sim \text{Gamma}(2, 1)$, which has prior mean two and allows strongly overdispersed components. For $p^{\mathrm{nb}}_k$, we use $b_\omega = 3$ -- the smallest integer value for which the implied prior on the odds $\omega_k$ has finite variance -- so that the centering rule in Eq.~(\ref{eq:centring}) simplifies to $a_\omega = (b_\omega - 1)(G-1)$. In the course of the data analysis in Sec.~\ref{sec:results}, we additionally explore the sensitivity of our results to all hyperparameter choices described above across a large grid of alternative specifications.

\subsection{Posterior Inference}
\label{sec:mcmc}

We explore the posterior distribution using a Gibbs sampler with data augmentation (\citealp{tanner1987calculation}). Each sampling iteration alternates between (i) imputing the latent data (the cluster membership of the observed data, as well as the size, cluster memberships, and marks of the hidden population) and (ii) updating model parameters given the observed data and the current imputations.

\textbf{Updating latent variables}. For each observed incident $i \in \mathcal{O}_g$, we draw its latent cluster label $z_i\in\{1,\dots,K\}$ from a categorical distribution with probabilities proportional to the product of the stratum-specific weight, the likelihood of the reporting pattern, and the likelihood of the mark:
\begin{equation}
P(z_i=k\mid\cdot)\propto \pi_{gk}
\left[\prod_{j=1}^R p_{kj}^{s_{ij}}(1-p_{kj})^{1-s_{ij}}\right]
f_{\mathrm{SNB}}(y_i\mid r_k, p^{\mathrm{nb}}_k),
\end{equation}
where $f_{\mathrm{SNB}}(\cdot)$ denotes the shifted negative binomial probability mass function implied by Eq.~(\ref{eq:snb}).

To account for incidents that are missed by all reporting systems, we augment the state space with an unobserved count $n_{0g}$ in each stratum. We first update the stratum intensity $\lambda_g$ from its Gamma full conditional,
\begin{equation}
\lambda_g \mid \cdot \sim \mathrm{Gamma}\!\left(m_g+n_{0g},\, 1\right),
\end{equation}
and then sample the number of missed incidents using Poisson thinning. Let
$p_{0g}=\sum_{k=1}^K \pi_{gk} q_k$ denote the marginal probability that an incident in stratum $g$ is missed entirely, where $q_k$ is the probability of the all-zero reporting pattern under cluster $k$. We draw
\begin{equation}
n_{0g}\mid\cdot \sim \mathrm{Poisson}(\lambda_g p_{0g}),
\end{equation}
and allocate the missed incidents to clusters via
\begin{equation}
(n_{0g1},\dots,n_{0gK}) \sim \mathrm{Multinomial}\!\left(n_{0g}, \frac{\pi_{g1}q_1}{p_{0g}},\dots,\frac{\pi_{gK}q_K}{p_{0g}}\right).
\end{equation}

We then impute fatality counts for the currently unobserved incidents assigned to cluster $k$ by drawing $n_{0gk}$ latent marks $y \sim f_{\mathrm{SNB}}(\cdot \mid r_k, p^{\mathrm{nb}}_k)$, and combine these draws with the observed $\{y_i:z_i=k\}$ to obtain a complete set of marks for cluster $k$. The imputed marks enter both the accumulation of the unobserved fatality total and the parameter updates below on an equal footing with the observed marks.

\textbf{Updating parameters.} Given the observed assignments and the augmented counts, all parameters except the concentration parameters $\alpha_g$ are updated using conjugate steps. Let $\mathcal{I}_k$ denote the set of indices for all incidents currently assigned to cluster $k$, comprising both the observed incidents and the indices of the augmented unobserved incidents in cluster $k$. The total number of incidents in cluster $k$ is $N_k = |\mathcal{I}_k|$.

The concentration parameter $\alpha_g$ has no closed-form full conditional and is updated with an adaptive Metropolis--Hastings step using normal proposals on the log scale. Writing $N_{gk}$ for the total number of incidents assigned to cluster $k$ in stratum $g$, the mixture weights are updated via
\begin{equation}
\boldsymbol{\pi}_g\mid\cdot \sim \mathrm{Dirichlet}\!\left(\frac{\alpha_g}{K}+N_{g1},\dots,\frac{\alpha_g}{K}+N_{gK}\right).
\end{equation}

Capture probabilities are shared across strata. Thus, each $p_{kj}$ is updated from its Beta full conditional using the complete set of assignments. Since unobserved incidents are by definition missed by all systems, we have $s_{ij}=0$ for all augmented incidents and the full conditional is 
\begin{equation}
p_{kj} \mid \cdot \sim \mathrm{Beta}\!\left(a_p + \sum_{i \in \mathcal{I}_k} s_{ij}, \quad b_p + \sum_{i \in \mathcal{I}_k} (1 - s_{ij})\right).
\end{equation}

The shifted negative binomial mark parameters $(r_k, p^{\mathrm{nb}}_k)$ are updated via a conjugate Gibbs step based on the Chinese restaurant table (CRT) data augmentation of \citet{zhou2015negative}. Writing $x_i = y_i - 1$ for the count component of the marks in cluster $k$, the probability parameter has the Beta full conditional
\begin{equation}
p^{\mathrm{nb}}_k \mid \cdot \sim \mathrm{Beta}\!\left(a_\omega + \sum_{i \in \mathcal{I}_k} x_i, \;\; b_\omega + N_k\, r_k\right).
\end{equation}
For the size parameter, we first draw the latent table counts $\ell_i \sim \mathrm{CRT}(x_i, r_k)$, which can be generated as a sum of independent Bernoulli variables, $\ell_i = \sum_{j=1}^{x_i} b_j$ with $b_j \sim \mathrm{Bernoulli}\big(r_k / (r_k + j - 1)\big)$ \citep{antoniak1974mixtures, zhou2015negative}. Conditional on $L_k = \sum_{i \in \mathcal{I}_k} \ell_i$, the size parameter is then updated from
\begin{equation}
r_k \mid \cdot \sim \mathrm{Gamma}\!\left(a_r + L_k, \;\; b_r - N_k \log(1 - p^{\mathrm{nb}}_k)\right),
\end{equation}
where the rate is positive because $-\log(1 - p^{\mathrm{nb}}_k) > 0$. Empty components ($N_k = 0$) simply draw $(r_k, p^{\mathrm{nb}}_k)$ from their priors.

Additional details on the MCMC algorithm are provided in the supplementary material (Sec.~\ref{sec:mcmc_details}), where we also present a simulation study (Sec.~\ref{sec:simulation}). In our simulation experiments, we evaluate the proposed model and sampler across a range of data-generating processes and compare performance to several competing approaches, from naive benchmark estimators to regression-based methods in the spirit of \citet{farcomeni2022many}, which we adapt to multi-list settings and to the shifted negative binomial likelihood in Sec.~\ref{sec:simulation}. The results show that our approach typically outperforms these approaches and yields accurate inference for both the total number of incidents and total fatalities, with well-calibrated posterior uncertainty. Empirical coverage of the 95\% credible intervals is close to nominal for the number of missed incidents (0.955 on average across three simulation settings, ranging from 0.935 to 0.965), whereas the intervals for the missed fatality total exhibit slight undercoverage (0.935 on average, ranging from 0.900 to 0.970), consistent with the mild re-use of the data through the empirical Bayes centering rule.

\section{Results}
\label{sec:results}

We estimate the model under four time stratifications and three list configurations. In the \textit{pooled} specification, all incidents form a single stratum ($G=1$). The \textit{year $\times$ season} specification splits each calendar year into a summer (April--September) and a winter (October--March) stratum ($G=24$). The \textit{annual} specification uses calendar years ($G=12$), while the \textit{monthly} specification uses calendar months pooled across years ($G=12$), isolating seasonal variation from long-run trends. Each stratification is combined with the three-, four-, and five-list configurations of Sec.~\ref{sec:lists}, yielding twelve specifications in total. 

Our preferred baseline specification uses \textit{four lists} with \textit{year $\times$ season strata}. The four-list configuration captures the main heterogeneity in reporting behavior without producing the sparse overlap cells of the five-list variant (see Supplementary Tab.~\ref{tab:mse_overlap_5L}). The year $\times$ season stratification is the finest temporal resolution at which all strata remain well populated, and it captures both seasonal and long run shifts in surveillance and crossing activity.

All presented results are based on 50,000 retained posterior draws per model run, obtained by running the sampler for 375,000 iterations, discarding an initial burn-in of 25,000 iterations, and thinning the remaining chain by a factor of seven. MCMC convergence is generally satisfactory; see the trace plots in Fig.~\ref{fig:trace} in the supplementary material.

\begin{figure}[!t]
    \centering

    \begin{subfigure}[b]{\textwidth}
        \centering
        \includegraphics[width=\textwidth, trim={0 0 0 0}, clip]{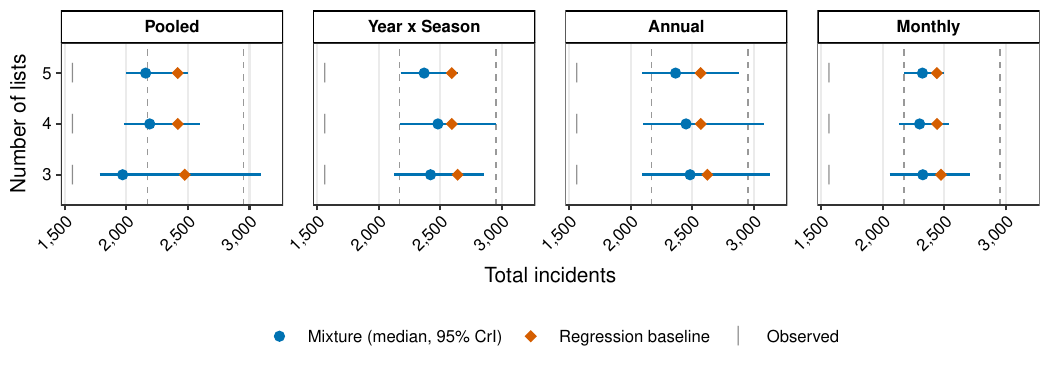}
        \caption{Total incidents.}
    \end{subfigure}\vspace{0.75em}
    \begin{subfigure}[b]{\textwidth}
        \centering
        \includegraphics[width=\textwidth, trim={0 0 0 0}, clip]{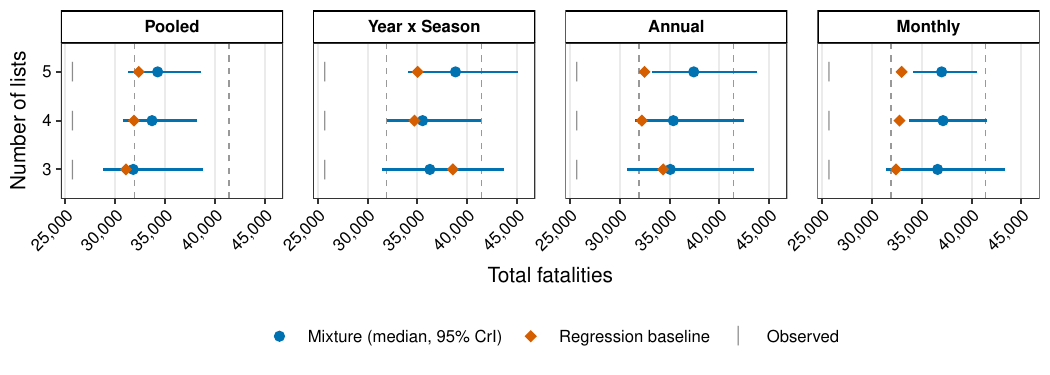}
        \caption{Total fatalities.}
    \end{subfigure}

    \caption{Posterior estimates of the total number of deadly incidents (top) and total fatalities (bottom) on the Central Mediterranean route, 2014--2025, across all list configurations (rows: three, four, and five lists) and time stratifications (panels: pooled, year $\times$ season, annual, monthly). Blue points and ranges are posterior medians and 95\% credible intervals of the Bayesian mixture model. Orange diamonds show the fully stratified, main-effects-only shifted negative binomial regression baseline. Grey vertical ticks mark the reported totals. The grey dashed rules indicate the 95\% credible interval of the baseline specification (four lists, year $\times$ season strata), extended across all panels for reference.}
    \label{fig:main_results}
\end{figure}

\textbf{Incident and Fatality Estimates.} Fig.~\ref{fig:main_results} summarizes our primary results for the estimated total number of incidents and total fatalities. Each panel corresponds to one time stratification and each row within a panel to one list configuration. Displaying all twelve combinations makes the sensitivity of the results to these a priori choices explicit. The observed totals (grey ticks) are identical across list configurations, since the three-, four-, and five-list settings re-aggregate the same underlying sources and every observed incident remains observed under each aggregation. Under the baseline specification, the 95\% credible interval for the total number of incidents spans 2,170--2,952 (posterior median 2,482), implying that about 53\%--72\% of all incidents are captured in the MMP data. The corresponding interval for total fatalities is 31,922--41,413 (posterior median 35,521). This translates to roughly 7--10 fatalities per day between February 17, 2014, and December 30, 2025 (compared to about 6 per day reported), suggesting that the MMP data capture approximately 62\%--81\% of total fatalities. In Sec.~\ref{sec:mort_est} in the supplementary material, we combine these estimates with additional data and assumptions to derive mortality \textit{rates} based on these fatality \textit{counts}. Results suggest a likely range of up to 3.5\% mortality risk per crossing attempt.

The vast majority of the mixture posteriors across alternative stratifications and list configurations concentrate within or very close to the baseline credible interval (grey dashed rules). One notable exception is the fully stratified five-list year $\times$ season setting for total fatalities, which yields a slightly higher range (34,060--45,075). However, this specification is quite sparse -- splitting the official list under the finest stratification leaves several capture patterns represented by only one or two observations -- so its estimates hinge on a small number of overlapping records. A dedicated prior sensitivity study, presented in Sec.~\ref{sec:prior-sensitivity} in the supplementary material, further indicates that the baseline mixture results are representative across 243 alternative prior specifications.

\begin{figure}[t]
    \centering
        \begin{subfigure}[b]{0.5\textwidth}
        \centering
        \includegraphics[width=\textwidth, trim={0 0 0 0}, clip]{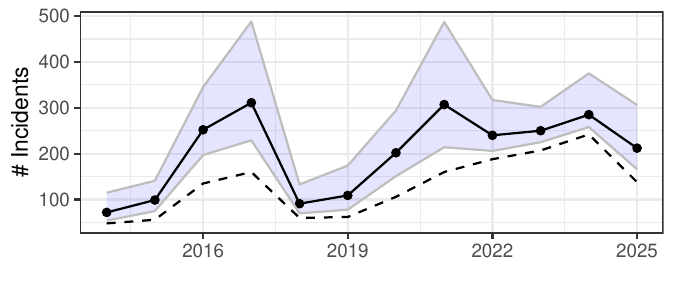}
        \caption{Total incidents by year.}
    \end{subfigure}%
    \begin{subfigure}[b]{0.5\textwidth}
        \centering
        \includegraphics[width=\textwidth, trim={0 0 0 0}, clip]{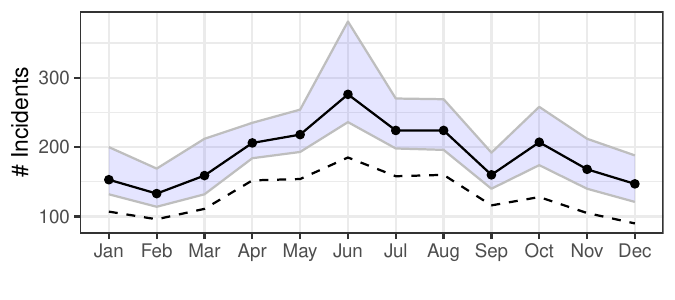}
        \caption{Total incidents by month.}
    \end{subfigure}\\
    \begin{subfigure}[b]{0.5\textwidth}
        \includegraphics[width=\textwidth, trim={0 0 0 0}, clip]{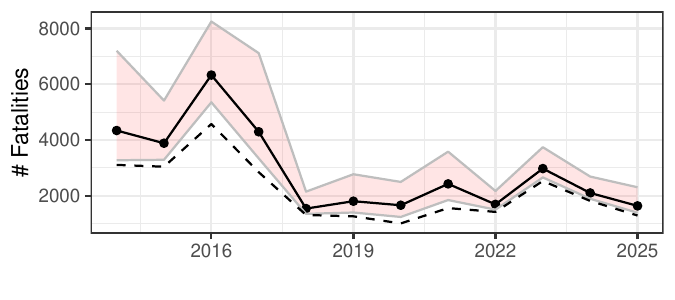}
        \caption{Total fatalities by year.}
    \end{subfigure}%
    \begin{subfigure}[b]{0.5\textwidth}
        \centering
        \includegraphics[width=\textwidth, trim={0 0 0 0}, clip]{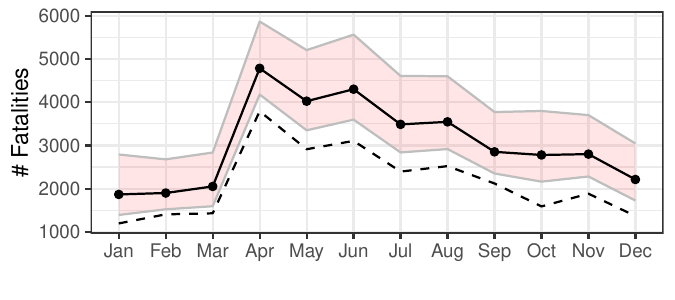}
        \caption{Total fatalities by month.}
    \end{subfigure}
    \caption{Posterior estimates of total incidents (top) and total fatalities (bottom), stratified by year (left) and month (right). Annual estimates are derived from the baseline year $\times$ season run by aggregating the per-stratum posterior draws over the two seasons of each year within every draw. Shaded areas are 95\% credible intervals. Solid black lines are posterior medians. Dashed black lines are reported totals.}
    \label{fig:res_strat}
\end{figure}

As a reference point from a second modeling paradigm, each cell of Fig.~\ref{fig:main_results} also reports a simple regression baseline (orange diamonds) in the spirit of \citet{farcomeni2022many}, see Sec.~\ref{sec:simulation} in the supplementary material for details. This baseline assumes no interactions between lists within strata -- an assumption we consider unlikely to hold exactly, but which provides a transparent and robust reference. Richer regression specifications allowing for list interactions regularly diverged in this application, as maximum likelihood estimates lack stability in the sparse within-stratum capture-pattern tables. Because list dependence is plausibly positive in this setting, neglecting it overstates surveillance coverage and biases the hidden population estimates downward. In line with this, the baseline estimates fall within or near the corresponding mixture credible intervals throughout, typically in the lower part of the interval.


\textbf{Time-Stratified Analysis.} Fig.~\ref{fig:res_strat} displays fatality and incident estimates stratified by year and by month. The annual estimates are obtained from the baseline year $\times$ season run by summing the two seasonal strata of each year within every posterior draw, which preserves the posterior dependence between seasons.\footnote{Note that 2014 does not cover a full calendar year: the incident record begins on February 17, 2014, so roughly one and a half months are missing, and 2014 totals should be read accordingly.} The monthly estimates are based on model runs stratified by calendar month, pooling months across years. Both levels of stratification reveal meaningful temporal variation. We find that fatalities were concentrated in the early years of the sample, especially prior to 2018. However, estimation uncertainty is also highest during these periods. This likely reflects the generally lower information content of the data during this period (see Sec.~\ref{sec:data}), together with the fact that fatality patterns become more stable after 2017, enabling greater information sharing across strata. Regarding seasonality, the majority of fatalities occur during the summer months. This aligns with peak crossing activity during this time of year due to favorable weather conditions (\citealp{zens2026dynamic}). 

\begin{figure}[!tbp]
    \centering

    \begin{subfigure}[b]{0.5\textwidth}
        \centering
        \includegraphics[width=\textwidth, trim={0 0 0 0}, clip]{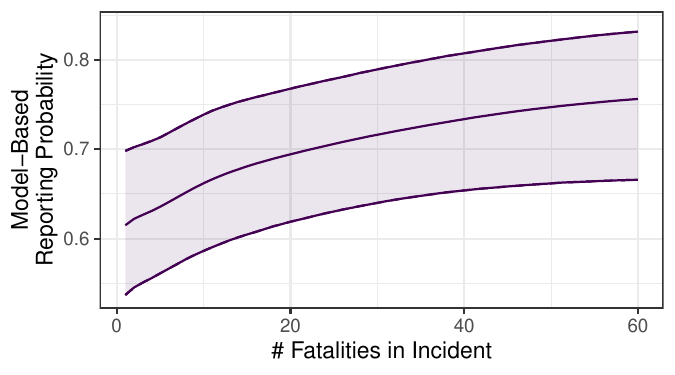}
        \caption{Probability of reporting by fatalities.}
    \end{subfigure}%
    \begin{subfigure}[b]{0.5\textwidth}
        \centering
        \includegraphics[width=\textwidth, trim={0 0 0 0}, clip]{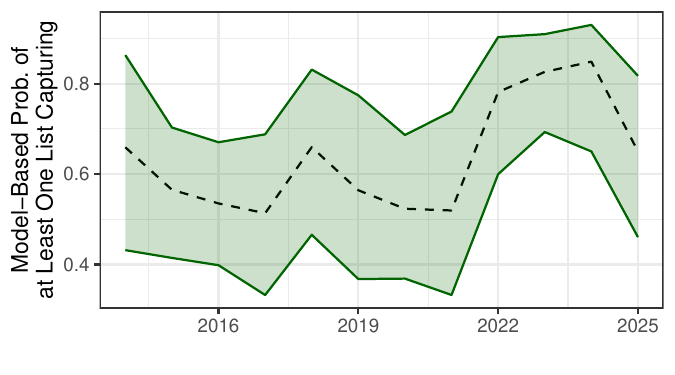}
        \caption{Probability of reporting over time.}
    \end{subfigure}

    \caption{Model-implied reporting probabilities. The left panel shows the relationship between the number of fatalities per incident and the probability of reporting. The right panel displays the probability of an incident being captured by at least one list over time. In both panels, the vertical axis represents the reporting probability. Lines denote posterior medians and shaded areas indicate 95\% credible intervals.}
    \label{fig:mark_vs_reporting}
\end{figure}

\textbf{Results on Reporting Patterns and Incident Severity.} Panel (a) in Fig.~\ref{fig:mark_vs_reporting} illustrates the model-implied relationship between the number of fatalities per incident and the probability of reporting. These estimates are derived by evaluating the model-implied marginal reporting probability for specific fatality counts at each MCMC draw. We estimate a positive correlation between incident severity and reporting probability. This confirms intuitive expectations regarding reporting patterns and underscores the necessity of jointly modeling reporting probabilities and fatalities in this context. Specifically, the posterior median reporting probability for incidents with few fatalities is slightly above 60\%, whereas it increases to roughly 75\% for larger incidents. Consequently, the undetected incidents tend to be smaller-scale with fewer fatalities. Interestingly, the model-implied probability of an incident being reported by at least one list remains relatively stable over time (panel (b) of Fig.~\ref{fig:mark_vs_reporting}). However, further exploratory analyses suggest this aggregate stability may mask two diverging trends -- an increasing probability of detecting large-scale events in recent years (Fig.~\ref{fig:fatality_reporting_time}), counterbalanced by a steady decline in the average fatality count per incident (Fig.~\ref{fig:incidents_fatalities_over_time}).

We additionally investigate the correlation structure between reporting and log fatalities using the fully augmented dataset (i.e., including imputed values for unobserved incidents). The resulting correlations (Fig.~\ref{fig:correlation}) corroborate the positive relationship between fatalities and reporting likelihood, and reveal intricate dependencies among the four reporting lists. Ultimately, these complex empirical patterns validate the need for a flexible, joint modeling framework as proposed in this article.


\section{Summary, Discussion, and Concluding Remarks}
\label{sec:conclusion} 

This article presents a probabilistic framework for estimating migrant fatalities en route, with a specific focus on Central Mediterranean crossings. We propose a novel methodology for marked capture-recapture problems, by extending Bayesian latent class analysis to accommodate auxiliary mark data via suitably chosen clustering kernels. We report three main findings. First, while recorded data successfully capture the majority of the mortality burden -- approximately 62\% to 81\% of total fatalities and 53\% to 72\% of fatal incidents -- a substantial share remains undocumented. Specifically, 95\% credible intervals indicate approximately 6,210--15,701 unobserved fatalities and 608--1,390 unobserved incidents. Second, regarding temporal trends, the absolute number of unreported fatalities appears to be concentrated in the 2014--2016 period. Although estimated aggregate underreporting shares remain relatively stable over time, this aggregate stability may mask more complex underlying dynamics, whereby reporting probabilities for large incidents increase while average incident size declines. Third, we find that the marginal reporting probability increases with incident severity, reflecting plausible real-world reporting behavior and underscoring the necessity of jointly modeling fatalities and reporting mechanisms in this domain.

\textbf{Limitations.} This study is subject to several limitations. First, as with any multiple-systems analysis method, the proposed model relies on fundamentally untestable assumptions, may suffer from identification issues, and may be sensitive to various prior choices (\citealp{manrique2013multiple}; \citealp{aleshin2024central}). We try to mitigate these concerns using estimates based on alternative modeling frameworks, alternative list constructions and stratifications, and prior sensitivity studies as robustness checks. Furthermore, the assumption of local independence within latent classes is an approximation. While latent classes absorb broad, common-cause drivers of co-reporting (e.g., incident severity, weather), some residual dependence between reporting channels may remain. If positive (e.g., media echoes other sources within latent classes), this dependence causes the model to overstate surveillance coverage, yielding conservative, downward-biased estimates for both unobserved incidents and total fatalities.

Second, the construction of the \textit{meta-observer} database via recoding MMP-reported sources into distinct lists may introduce measurement error and bias. A primary concern here is incomplete recording of sources that did in fact report an incident. For example, if the MMP database systematically records only the first or first few sources associated with an event, observed overlap between lists may be biased downwards. In standard multiple-systems settings with positive list dependence, this tends to increase estimated coverage and thereby reduce the estimated hidden population, yielding conservative estimates of unobserved incidents. If this mismeasurement were unrelated to incident severity, the same logic would imply conservative estimates of total fatalities. In the present setting, however, the direction of bias for fatality totals is not guaranteed, because source-recording errors may vary systematically with incident size and with the specific combination of lists involved. While our aggregation strategy using rather broad lists mitigates some of these issues -- and while the three- and five-list re-estimations probe sensitivity to the recoding directly -- it ultimately does not safeguard our analyses from such biases. 

Third, our study takes MMP-reported fatalities at face value. However, source-specific fatality counts can vary, especially for high-fatality incidents reliant on eyewitness accounts. The MMP reports the most conservative verified number without recording alternatives. This introduces a downward bias in the observed marks, implying that the presented total fatality estimates are lower bounds. Relatedly, our results are conditional on the scope and method of the underlying data collection. The estimand is thus the count of incidents and fatalities undocumented by the IOM MMP system. Alternative data collection efforts might produce substantially different baseline counts and estimates such as ours therefore continue to carry uncertainty that is tied to the data collection process itself.

Finally, nuanced definitional issues may arise in case different sources do not define incidents and fatalities consistently over time, potentially introducing further bias. Such inconsistencies are unlikely to be substantial given the relative clarity of the target population and the unambiguous nature of death or missingness on the open sea. Nevertheless, they cannot be fully resolved without a qualitative investigation into every cited source. Future audits of the raw data and interviews with source representatives could help quantify the potential extent of this issue.

\textbf{Future Research.} Future methodological research may address some of these concerns by incorporating an additional measurement error layer into the statistical model; for instance, along the lines of joint multiple systems estimation and record linkage models (\citealp{fienberg2009integrated}; \citealp{tancredi2011hierarchical}). However, identification may remain challenging in the absence of a `gold standard' list. Alternatively, a retrospective full source audit of captured incidents could be used to improve the robustness of the MMP data. This would, however, be a highly resource-intensive exercise beyond the scope of this article.

Developing an objective or reference prior framework for this model class would also be beneficial, given the typical sensitivity of capture-recapture models to prior choices. While we addressed this through sensitivity analyses, a more principled solution regarding default priors would be a valuable theoretical contribution.


From a modeling perspective, extending the framework to incorporate smooth temporal dynamics offers a valuable refinement. Because our latent class parameters are currently shared across strata, the model inherently approximates any structural improvements in surveillance capacity as compositional shifts in incident archetypes. While this shared-parameter design ensures estimation stability, future work could allow capture probabilities to evolve continuously over time to better isolate systemic changes in reporting. Furthermore, we leave for future work replacing the simple Poisson model for the incident process $N_g$ with covariate-driven Poisson or Poisson--log-normal intensity specifications, which would allow the latent number of incidents to depend smoothly on time or on external drivers such as crossing activity and weather. However, such extensions would come at the expense of more complex posterior inference. Another promising pathway is to develop flexible regression-based multiple-systems approaches with time-varying parameters. This could enhance stability of maximum likelihood estimates in sparse stratified settings where fully separate per-stratum fits are fragile, as we experienced in our setting.

More broadly, working towards a more comprehensive and robust data environment is essential for understanding migrant mortality and informing policy. 
Here, an important future research effort is to merge information contained in the MMP with auxiliary information in related databases.\footnote{For instance, the \textit{UNITED List of Refugee Deaths} used in \citet{farcomeni2022many} is an alternative, media-focused migrant death list that provides data from 1993--2025.} However, such matching exercises are far from straightforward, require extensive efforts and are beyond the scope of this study. Overall, we view this paper primarily as a methodological contribution that future estimation efforts can build on.

In conclusion, while our estimates remain contingent on incomplete data and modeling assumptions, they provide a principled and uncertainty-aware alternative to naive adjustment rules and help clarify the likely magnitude of structural undercount. Ultimately, this work offers a foundation for improved mortality surveillance in migration contexts and paves the way for policy evaluation and monitoring. Importantly, the utility of this framework extends beyond the Central Mediterranean. It is readily applicable to other migration routes and, more broadly, to related capture-recapture settings providing auxiliary mark data, such as estimating civilian casualties in conflict or ecological surveys recording animal size.

\begin{acks}[Acknowledgments]
The authors are grateful to Alessio Farcomeni, Anita Gohdes, Gertraud Malsiner-Walli, the \textit{Missing Migrants Project} team at IOM, and participants of the \textit{Vienna Workshop on Migration Modeling 2026} and of the research seminar of the Faculty of Economics and Statistics at the University of Innsbruck for insightful discussions and comments that helped shape this paper. The authors declare that they have no conflict of interest.
\end{acks}

\begin{funding}
Gregor Zens received financial support from the European Research Council (ERC) through the \textit{2C-RISK} grant (Grant Agreement No. 101162653). Zoe Sigman received funding from the Deutsche Forschungsgemeinschaft (DFG, German Research Foundation) -- 390285477/GRK 2458. Views and opinions expressed are, however, those of the author(s) only and do not necessarily reflect those of the European Union or the European Research Council Executive Agency. Neither the European Union nor the granting authority can be held responsible for them.
\end{funding}

\subsection*{Data Availability Statement}\label{data-availability-statement}

The \textit{Missing Migrants Project} data are publicly available online at \url{https://missingmigrants.iom.int/}. Data on attempted crossings are available online at \url{https://data.humdata.org/dataset/iom-missing-migrants-project-data}. The recoded dataset and the estimation and replication code are available from the authors upon request.

\singlespacing
\bibliographystyle{apalike} 
\bibliography{lit}       


\newpage
\setcounter{equation}{0}
\setcounter{figure}{0}
\setcounter{section}{0}
\setcounter{table}{0}
\renewcommand\thesection{S\arabic{section}}
\renewcommand{\theHsection}{S\arabic{section}}
\renewcommand\theequation{S\arabic{equation}}
\renewcommand\thefigure{S\arabic{figure}}
\renewcommand\thetable{S\arabic{table}}

\begin{center}
    
\vspace*{2em}\noindent{\large\MakeUppercase{{Supplementary Material}}}\vspace*{1em}

{\large\textbf{Probabilistic Estimation of Hidden Migrant Fatalities\\[4pt] Along the Central Mediterranean Route}}\\[1.5em]
{\normalsize Gregor Zens$^{1}$ and Zoe Sigman$^{2}$}\\[1em]
{\small $^{1}$International Institute for Applied Systems Analysis (IIASA)\\[0.5em]
$^{2}$DYNAMICS, The Hertie School and Humboldt University of Berlin}\\[1em]
\end{center}


The supplementary material contains further MCMC details and derivations, a simulation study, a prior sensitivity analysis, an exploratory analysis of migrant mortality rates, and additional figures and tables.

\section{Further MCMC Details and Derivations}
\label{sec:mcmc_details}

In this section, we provide further details on the Gibbs sampling steps used for posterior inference. To facilitate estimation, we employ a data augmentation scheme that includes the latent cluster assignments $\mathbf{z}$, the number of unobserved incidents $\mathbf{n}_0 = \{n_{0g}\}_{g=1}^G$, and the latent marks for these unobserved incidents. Let the collection of all model parameters be $\boldsymbol{\Theta} = \{\boldsymbol{\pi}, \mathbf{p}, \mathbf{r}, \mathbf{p}^{\mathrm{nb}}, \boldsymbol{\lambda}, \boldsymbol{\alpha}\}$. For a specific stratum $g$, let $m_g$ denote the number of observed incidents and $n_{0g}$ the number of unobserved incidents. The observed data consist of reporting patterns $\mathbf{s}_i$ and integer marks $y_i$ for $i \in \mathcal{O}_g$. The unobserved incidents are characterized by latent marks $\mathbf{y}_{0g} = \{y_{0g\ell}\}_{\ell=1}^{n_{0g}}$ and are missed by all reporters, i.e. each unobserved incident has the all-zero capture pattern.

The joint probability of the data and parameters factors into an observed component and an unobserved component. Denoting the total stratum count as $N_g = m_g + n_{0g}$, the joint likelihood is proportional to
\begin{align}
    P(\text{Data}, \boldsymbol{\Theta}) \propto \;& P(\boldsymbol{\Theta}) \times \prod_{g=1}^G \left[ \frac{e^{-\lambda_g} \lambda_g^{N_g}}{N_g!} \times L_{\text{obs}, g} \times L_{\text{miss}, g} \right],
\end{align}
where $P(\boldsymbol{\Theta})$ denotes the joint prior. The contribution from the $m_g$ observed incidents is
\begin{equation}
    L_{\text{obs}, g} = \prod_{i=1}^{m_g} \left( \sum_{k=1}^K \pi_{gk} \, f_{\mathrm{SNB}}(y_i \mid r_{k}, p^{\mathrm{nb}}_{k}) \prod_{j=1}^R p_{kj}^{s_{ij}} (1-p_{kj})^{1-s_{ij}} \right),
\end{equation}
and the contribution from the $n_{0g}$ unobserved incidents is
\begin{equation}
L_{\text{miss}, g} = \prod_{\ell=1}^{n_{0g}} \left( \sum_{k=1}^K \pi_{gk} \, f_{\mathrm{SNB}}(y_{0g\ell} \mid r_{k}, p^{\mathrm{nb}}_{k}) \prod_{j=1}^R (1-p_{kj}) \right).
\end{equation}
Note that the unobserved term simplifies due to the all-zero reporting patterns for unobserved incidents. Let the non-detection probability of an incident in cluster $k$ be $q_k = \prod_{j=1}^R (1-p_{kj})$.

\subsection{Augmenting Missing Observations}

To sample the missing observations, we utilize the properties of a Poisson process under random thinning. We update the unobserved state by sampling from the joint posterior of the tuple $\{n_{0g}, \mathbf{z}_{0g}, \mathbf{y}_{0g}\}$. Conditional on the parameters $\boldsymbol{\Theta}$, this joint density is proportional to the unobserved component of the likelihood. Explicitly, we have:
\begin{equation}
P(n_{0g}, \mathbf{z}_{0g}, \mathbf{y}_{0g} \mid  \boldsymbol{\Theta})
\propto
\frac{\lambda_g^{n_{0g}} e^{-\lambda_g p_{0g}}}{n_{0g}!}
\prod_{\ell=1}^{n_{0g}}
\left(
\pi_{g, z_{0g\ell}} q_{z_{0g\ell}}
f_{\mathrm{SNB}}(y_{0g\ell} \mid r_{z_{0g\ell}}, p^{\mathrm{nb}}_{z_{0g\ell}})
\right).
\end{equation}
To sample from this joint distribution efficiently, we factorize it into a marginal distribution for the count $n_{0g}$ and a conditional distribution for the characteristics $(\mathbf{z}_{0g}, \mathbf{y}_{0g})$ given $n_{0g}$. We can rewrite the term inside the product by multiplying and dividing by the marginal missing probability $p_{0g} = \sum_{k=1}^K \pi_{gk} q_k$:
\begin{equation}
    \pi_{g, z_{0g\ell}} q_{z_{0g\ell}} f_{\mathrm{SNB}}(y_{0g\ell} \mid r_{z_{0g\ell}}, p^{\mathrm{nb}}_{z_{0g\ell}}) = p_{0g} \times \underbrace{\frac{\pi_{g, z_{0g\ell}} q_{z_{0g\ell}}}{p_{0g}}}_{P(z_{0g\ell} \mid \mathbf{s}=\mathbf{0})} \times f_{\mathrm{SNB}}(y_{0g\ell} \mid r_{z_{0g\ell}}, p^{\mathrm{nb}}_{z_{0g\ell}}).
\end{equation}
Substituting this factorization back into the joint density expression yields:
\begin{align}
    P(n_{0g}, \mathbf{z}_{0g}, \mathbf{y}_{0g} \mid \cdot) &\propto \frac{\lambda_g^{n_{0g}} e^{-\lambda_g p_{0g}}}{n_{0g}!} \prod_{\ell=1}^{n_{0g}} \left( p_{0g} \cdot P(z_{0g\ell} \mid \mathbf{s}=\mathbf{0}) \cdot f_{\mathrm{SNB}}(y_{0g\ell} \mid r_{z_{0g\ell}}, p^{\mathrm{nb}}_{z_{0g\ell}}) \right) \\
    &= \underbrace{\frac{(\lambda_g p_{0g})^{n_{0g}} e^{-\lambda_g p_{0g}}}{n_{0g}!}}_{\text{Poisson PMF for } n_{0g}} \times \prod_{\ell=1}^{n_{0g}} \underbrace{\left( P(z_{0g\ell} \mid \mathbf{s}=\mathbf{0}) \cdot f_{\mathrm{SNB}}(y_{0g\ell} \mid r_{z_{0g\ell}}, p^{\mathrm{nb}}_{z_{0g\ell}}) \right)}_{\text{i.i.d. draws for } z, y}.
\end{align}
This factorization justifies the following sequential sampling strategy:
\begin{enumerate}
    \item \textbf{Update intensity.} First, update $\lambda_g$ given the total count $N_g = m_g + n_{0g}$. Under the improper scale prior $p(\lambda_g) \propto \lambda_g^{-1}$, the full conditional is
    \begin{equation}
        \lambda_g \sim \text{Gamma}(N_g, 1),
    \end{equation}
    which is proper whenever $N_g \geq 1$.
    \item \textbf{Sample marginal count.} Draw the new number of unobserved incidents:
    \begin{equation}
        n_{0g} \sim \text{Poisson}(\lambda_g p_{0g}).
    \end{equation}
    \item \textbf{Sample conditional characteristics.} For each of the $n_{0g}$ incidents, draw a cluster assignment $z \sim \text{Categorical}(\mathbf{w}^{\text{miss}}_g)$, where the normalized weights are $w_{gk}^{\text{miss}} = \frac{\pi_{gk} q_k}{p_{0g}}$. To arrive at the complete data used to update the model parameters, then draw a mark $y \sim f_{\mathrm{SNB}}(\cdot \mid r_z, p^{\mathrm{nb}}_z)$ for each of the $n_{0g}$ incidents.
\end{enumerate}

\subsection{CRT Data Augmentation for the Shifted Negative Binomial Marks}
\label{sec:crt}

The mark parameters $(r_k, p^{\mathrm{nb}}_k)$ are updated on the augmented data of each component, i.e., the observed marks together with the imputed marks of the missing incidents currently allocated to component $k$. Writing $x_i = y_i - 1$ for the count component and $N_k$ for the total (observed plus augmented) number of incidents in cluster $k$, the negative binomial likelihood is conjugate to the Beta prior in $p^{\mathrm{nb}}_k$ for fixed $r_k$, yielding
\begin{equation}
p^{\mathrm{nb}}_k \mid \cdot \sim \mathrm{Beta}\!\left(a_\omega + \sum_{i \in \mathcal{I}_k} x_i, \; b_\omega + N_k r_k\right).
\end{equation}

For the size parameter $r_k$, we use the Chinese restaurant table (CRT) augmentation of \citet{zhou2012augment} and \citet{zhou2015negative}. A negative binomial variable $x \sim \mathrm{NB}(r, p)$ admits the compound-Poisson representation $x = \sum_{t=1}^{\ell} u_t$, where $\ell \sim \mathrm{Poisson}\!\left(-r \log(1-p)\right)$ and the $u_t$ are i.i.d.\ logarithmic random variables with parameter $p$. Conditional on $x$ and $r$, the latent count $\ell$ follows the CRT distribution, $\ell \sim \mathrm{CRT}(x, r)$, which is rooted in the P\'olya-urn representation of the Dirichlet process \citep{antoniak1974mixtures} and can be generated as a sum of independent Bernoulli variables,
\begin{equation}
\ell = \sum_{j=1}^{x} b_j, \qquad b_j \sim \mathrm{Bernoulli}\!\left(\frac{r}{r + j - 1}\right).
\end{equation}
Given the table counts $\ell_i \sim \mathrm{CRT}(x_i, r_k)$ for all $i \in \mathcal{I}_k$ and their sum $L_k = \sum_{i \in \mathcal{I}_k} \ell_i$, the conditional likelihood of $r_k$ is proportional to $r_k^{L_k} \exp\{ r_k N_k \log(1 - p^{\mathrm{nb}}_k)\}$, which combines with the $\mathrm{Gamma}(a_r, b_r)$ prior to the conjugate update
\begin{equation}
r_k \mid \cdot \sim \mathrm{Gamma}\!\left(a_r + L_k, \; b_r - N_k \log(1 - p^{\mathrm{nb}}_k)\right).
\end{equation}
The rate parameter is strictly positive since $-\log(1 - p^{\mathrm{nb}}_k) > 0$. Components with no observed or imputed marks draw $(r_k, p^{\mathrm{nb}}_k)$ from their priors. 

\subsection{Updating Mixture Weights and Sparsity Parameter}

Let $N_{gk}$ denote the total number of incidents (both observed and augmented) assigned to cluster $k$ in stratum $g$. To improve mixing, we update the concentration parameter $\alpha_g$ while marginalizing out the weights $\boldsymbol{\pi}_g$. The marginal likelihood of the cluster counts $\mathbf{N}_g = (N_{g1}, \dots, N_{gK})$ under the Dirichlet-Multinomial model is:
\begin{equation}
    P(\mathbf{N}_g \mid \alpha_g) \propto \frac{\Gamma(\alpha_g)}{\Gamma(\alpha_g + N_g)} \prod_{k=1}^K \frac{\Gamma(\alpha_g/K + N_{gk})}{\Gamma(\alpha_g/K)}.
\end{equation}
We update $\alpha_g$ via a Metropolis-Hastings step using a random walk proposal on the log-scale, $\log(\alpha_g^*) \sim \mathcal{N}(\log(\alpha_g), \tau^2)$, where $\tau^2$ is a tuning parameter adapted to achieve acceptance rates around $0.234$. The acceptance probability for the proposal $\alpha_g^*$ is $\min(1, r)$, where
\begin{equation}
    r = \frac{P(\mathbf{N}_g \mid \alpha_g^*) \, p(\alpha_g^*)}{P(\mathbf{N}_g \mid \alpha_g) \, p(\alpha_g)} \times \frac{\alpha_g^*}{\alpha_g},
\end{equation}
with $p(\cdot)$ denoting the Gamma prior density. The ratio $\alpha_g^*/\alpha_g$ is the Jacobian correction required for the log-scale proposal. Conditional on the current $\alpha_g$, we then update the weights:
\begin{equation}
    \boldsymbol{\pi}_g \mid \dots \sim \text{Dirichlet}\left( \frac{\alpha_g}{K} + N_{g1}, \dots, \frac{\alpha_g}{K} + N_{gK} \right).
\end{equation}

\section{Simulation Study}
\label{sec:simulation}

\subsection{Data Generating Processes}

We study three data-generating settings for a closed population of size $N=2500$ observed on $R=4$ capture lists, using 200 Monte Carlo replicates per setting. All data-generating processes use shifted negative binomial marks, $y_i = 1 + x_i$ with $x_i \sim \mathrm{NB}(r, p^{\mathrm{nb}})$ in the parameterization of Eq.~(\ref{eq:snb}) in the main text. All Bayesian results are based on 50,000 retained posterior draws, obtained by running the sampler for 375,000 iterations, discarding an initial burn-in of 25,000 iterations, and thinning the remaining chain by a factor of seven, using the default specification described in Sec.~\ref{sec:hyperparameters} of the main text.

\textbf{Setting A: Independence.} In the simplest setup we consider, for each replicate we generate marks $y_i = 1 + x_i$ with $x_i \sim \mathrm{NB}(r, p^{\mathrm{nb}})$, $r = 10$ and $p^{\mathrm{nb}} = 2/3$, implying a mean mark of $1 + r\,p^{\mathrm{nb}}/(1-p^{\mathrm{nb}}) = 21$ with substantial overdispersion. Capture indicators are then generated independently across incidents and lists as $s_{ij}\sim\mathrm{Bernoulli}(p_j)$ with $\mathbf{p}=(0.40,0.10,0.12,0.20)$ for $j=1,\ldots,4$. Thus there is no relationship between $y_i$ and capture, and there are no list interactions.

\textbf{Setting B: Mixture (latent classes).} In each replicate we generate a three-class mixture with weights $\boldsymbol{\pi}=(0.40,0.30,0.30)$, and class memberships are sampled from $z_i\sim\mathrm{Categorical}(\boldsymbol{\pi})$ for $i=1,\ldots,N$. Conditional on $z_i=k$, marks are drawn from shifted negative binomial distributions with size parameters $\mathbf{r}=(20,10,5)$ and probability parameters chosen such that the class-specific mean marks equal $1 + (500, 80, 10)$. Given $z_i$, captures are independent across lists with $s_{ij}\mid(z_i=k)\sim\mathrm{Bernoulli}(p_{kj})$, where
\[
\begin{pmatrix}
\mathbf{p}_1\\
\mathbf{p}_2\\
\mathbf{p}_3
\end{pmatrix}
=
\begin{pmatrix}
0.60 & 0.55 & 0.60 & 0.55\\
0.35 & 0.30 & 0.35 & 0.30\\
0.15 & 0.18 & 0.15 & 0.18
\end{pmatrix}.
\]
The third class has both smaller marks and lower capture probabilities, inducing a positive association between severity and detectability, while the large-mark class is captured at high rates.

\textbf{Setting C: Graded detectability.} In each replicate we generate a three-class mixture with weights $\boldsymbol{\pi}=(0.40,0.35,0.25)$ and assignments $z_i\sim\mathrm{Categorical}(\boldsymbol{\pi})$. Conditional on $z_i=k$, marks follow shifted negative binomial distributions with size parameters $\mathbf{r}=(15,8,5)$ and class-specific mean marks $1 + (60, 25, 8)$. Captures are conditionally independent given $z_i$ with
\[
\begin{pmatrix}
\mathbf{p}_1\\
\mathbf{p}_2\\
\mathbf{p}_3
\end{pmatrix}
=
\begin{pmatrix}
0.45 & 0.42 & 0.45 & 0.42\\
0.35 & 0.33 & 0.35 & 0.33\\
0.28 & 0.26 & 0.28 & 0.26
\end{pmatrix}.
\]
This setting is more demanding than Setting B, as overall detection is lower and the large-mark class is only moderately detected, so a genuine share of the unobserved fatality mass stems from large incidents.

\subsection{Competitor Frameworks}

All competitor methods, including the Bayesian mixture model, are fit to the observed (zero-truncated) sample of size $m$, with observed marks $y_1,\ldots,y_m$ and capture matrix $\mathbf{S}\in\{0,1\}^{m\times R}$. Conceptually, all competitor models use the same large-sample decomposition suggested by \citet{farcomeni2022many}, where the unobserved total marks can be approximated by the number of missed units times the average mark among missed units. Let $n_0=\#\{i:\mathbf{s}_i=\mathbf{0}\}$ be the number of unobserved units, let $D_0=\sum_{i:\mathbf{s}_i=\mathbf{0}} y_i$ be the corresponding unobserved total, and let $d_0=\mathbb{E}[y_i \mid \mathbf{s}_i=\mathbf{0}]$ denote the mean mark for missed units. Then, as $n_0$ grows, the sample average mark among missed units concentrates around $d_0$, so that $D_0 \approx n_0 d_0$ becomes a useful approximation; see \citet{farcomeni2022many} for simulation evidence. In practice, this estimator is implemented by separately estimating $\widehat{n}_0$ from the capture histories and $\widehat{d}_0$ from a mark model, and reporting the plug-in estimate $\widehat{D}_0=\widehat{n}_0\,\widehat{d}_0$. We summarize the benchmark models we use as follows.

\subsubsection{Naive estimator (Chao)} Let $c_i=\sum_{j=1}^R s_{ij}$ denote the number of lists on which observed unit $i$ appears, and define $f_1=\#\{i:c_i=1\}$ and $f_2=\#\{i:c_i=2\}$. The naive estimate of the number of missed units, based on \citet{chao1987}, is
\[
\widehat{n}_0^{\text{naive}}=\frac{f_1^2}{2f_2}.
\]
The mean mark among missed units is naively taken to equal the observed sample mean,
\[
\widehat{d}_0^{\text{naive}}=\frac{1}{m}\sum_{i=1}^m y_i,
\]
so the missed total can be estimated as
\[
\widehat{D}_0^{\text{naive}}=\widehat{n}_0^{\text{naive}}\ \widehat{d}_0^{\text{naive}}.
\]

\subsubsection{Fixed-order regression models.}
\label{sec:regression-competitors}

We consider first a regression-based procedure using two separate regressions; first, a model for marks to estimate $\widehat{d}_0$, and second, a log-linear Poisson model for capture-pattern counts to estimate $\widehat{n}_0$.

\textbf{Mark model.} To match the count nature of the marks -- and to place the maximum-likelihood competitors on the same likelihood footing as the Bayesian mixture -- the mark regression is a \textit{shifted negative binomial regression} with log-mean link \citep[cf.][]{zhou2012lognormal}:
\[
y_i = 1 + x_i, \qquad x_i \sim \mathrm{NB}(r, \mu_i), \qquad \log \mu_i = \beta_0 + \sum_{j=1}^R \beta_j s_{ij},
\]
where $\mathrm{NB}(r, \mu_i)$ denotes the negative binomial with size $r$ and mean $\mu_i$ (equivalently, $p^{\mathrm{nb}}_i = \mu_i / (\mu_i + r)$ in the parameterization of the main text), and $r$ is a common dispersion parameter estimated jointly with the coefficients by maximum likelihood. For a \textit{main effects only} model the linear predictor is as displayed. We also consider a version with all \textit{pairwise interactions},
\[
\log \mu_i = \beta_0 + \sum_{j=1}^R \beta_j s_{ij}
+\sum_{1\le j<\ell\le R}\beta_{j\ell} s_{ij}s_{i\ell}.
\]

The target is the expected mark for an all-zero capture history. Since all indicators (and hence all interaction products) equal zero when $s_{i1}=\cdots=s_{iR}=0$, the corresponding predictor reduces to $\widehat{\mu}_0 = \exp(\widehat{\beta}_0)$, and the implied mean mark is
\[
\widehat{d}_0 = 1 + \widehat{\mu}_0.
\]

\textbf{Event model.} Let $\mathbf{s}_i$ denote the capture pattern of observed unit $i$, and define pattern counts
\[
n(\mathbf{s})=\#\{i:\mathbf{s}_i=\mathbf{s}\},\qquad \mathbf{s}\in\{0,1\}^R.
\]
A Poisson GLM is fit to the $2^R-1=15$ observed cells, omitting the unobserved all-zero cell $\mathbf{s}=\mathbf{0}$. Again, we consider a \textit{main effects} only model
\[
n(\mathbf{s})\sim\mathrm{Poisson}(\lambda(\mathbf{s})),\qquad
\log\lambda(\mathbf{s})=\gamma_0+\sum_{j=1}^R \gamma_j s_j,
\]
and a model with all \textit{pairwise interactions}
\[
\log\lambda(\mathbf{s})=\gamma_0+\sum_{j=1}^R\gamma_j s_j+\sum_{1\le j<\ell\le R}\gamma_{j\ell}s_js_\ell.
\]

The estimate of the missing all-zero cell is obtained by prediction at $\mathbf{s}=\mathbf{0}$,
\[
\widehat{n}_0=\widehat{\lambda}(\mathbf{0})=\exp(\widehat{\gamma}_0),
\]
and the missed total is then
\[
\widehat{D}_0=\widehat{n}_0\ \widehat{d}_0.
\]

\subsubsection{Regression with model selection over hierarchical log-linear models}

A second regression competitor fits a \emph{family} of hierarchical models for both the marks regression and the incidents regression, and then selects models by AIC or BIC, computed separately for marks and incidents. For this, we treat the four lists as binary factors $L1,\ldots,L4$. Candidate specifications always include all four main effects and may include any subset of the six pairwise interactions $L1{:}L2,\ldots,L3{:}L4$, subject to a hierarchy constraint. In total, this yields $2^6 = 64$ possible hierarchical models. 

Otherwise, the mark and event models follow the same ideas as for the regression models outlined in Sec.~\ref{sec:regression-competitors}. For mark models and incident models separately, we compute AIC and BIC over the 64 candidates and select the minimizer under each criterion, yielding $(\widehat{d}_0^{\text{AIC}},\widehat{n}_0^{\text{AIC}})$ and $(\widehat{d}_0^{\text{BIC}},\widehat{n}_0^{\text{BIC}})$. The missed totals are reported as $\widehat{D}_0^{\text{BIC}}=\widehat{n}_0^{\text{BIC}}\widehat{d}_0^{\text{BIC}}$ and $\widehat{D}_0^{\text{AIC}}=\widehat{n}_0^{\text{AIC}}\widehat{d}_0^{\text{AIC}}$, respectively.

\subsection{Results}

Table~\ref{tab:simulation_results} summarizes the results of the simulation experiments in terms of average relative error (Panel A) and log root mean squared error (Panel B). Unsurprisingly, the naive estimator is strongly biased and exhibits the largest error across all settings. While regression methods perform well under independence, they incur substantial bias for $N_0$ and $D_0$ under the dependent data generating processes. This can be mitigated to some extent by employing model selection based on information criteria, consistent with considerations in \citet{silverman2020multiple}. The mark-augmented latent class model demonstrates high accuracy across all settings, exhibiting minimal bias and achieving the lowest log-RMSE overall for both the missed incident count and the missed fatality total. We note that none of the simulation settings involves time stratification, for which the optimal treatment within the regression paradigm is unclear. 

We also evaluate the calibration of the 95\% posterior credible intervals of the Bayesian model via empirical coverage across the 200 replicates per setting. For the number of missed incidents $N_0$, we observe empirical rates of 0.935, 0.965, and 0.965 in the independence, mixture, and graded settings, respectively (0.955 overall), which is close to the nominal level. For the missed mark sum $D_0$, we record empirical coverage of 0.935, 0.970, and 0.900 (0.935 overall), i.e., slight undercoverage, most visibly in the graded setting. A plausible contributing factor is the empirical Bayes centering of the mark prior on the observed marks (Sec.~\ref{sec:priors}), which mildly re-uses the data and can shade posterior uncertainty for the fatality total. Overall, however, coverage is quite decent -- particularly considering that these rates are computed from a finite number of replicates, so that some Monte Carlo noise is expected.

\begin{table}[t]
\footnotesize
\centering
\caption{Simulation Results: Bias and RMSE under three simulation settings}
\label{tab:simulation_results}
\begin{threeparttable}
\begin{tabular}{lrrrrrrrr}
\toprule
 & \multicolumn{2}{c}{Independence} 
 & \multicolumn{2}{c}{\makecell{Mixture \\ (Latent Class)}} 
 & \multicolumn{2}{c}{\makecell{Graded \\ Detectability}} 
 & \multicolumn{2}{c}{Overall} \\
\cmidrule(lr){2-3} \cmidrule(lr){4-5} \cmidrule(lr){6-7} \cmidrule(lr){8-9}
Model & $D_0$ & $N_0$ & $D_0$ & $N_0$ & $D_0$ & $N_0$ & $D_0$ & $N_0$ \\
\midrule
\multicolumn{9}{l}{\textit{Panel A: Average Relative Error}} \\
Naive & 0.665 & 0.666 & 2.782 & -0.194 & 0.719 & 0.242 & 1.389 & 0.238\\
Regression (Main Effects) & 0.000 & 0.002 & -0.289 & -0.598 & -0.068 & -0.119 & -0.119 & -0.238\\
Regression (Pairwise Interactions) & 0.021 & 0.013 & -0.155 & -0.142 & 0.002 & 0.005 & -0.044 & -0.042\\
Regression (AIC) & 0.011 & 0.008 & -0.131 & -0.145 & -0.035 & -0.074 & -0.052 & -0.070\\
Regression (BIC) & 0.001 & 0.003 & 0.388 & -0.150 & -0.033 & -0.086 & 0.119 & -0.078\\
Bayesian Mixture & -0.004 & -0.004 & 0.019 & -0.013 & 0.078 & 0.027 & 0.031 & 0.003\\
\midrule
\multicolumn{9}{l}{\textit{Panel B: Log RMSE}} \\
Naive & 9.511 & 6.468 & 11.415 & 4.761 & 9.076 & 4.750 & 10.881 & 5.950\\
Regression (Main Effects) & 7.382 & 4.322 & 9.240 & 5.807 & 7.167 & 4.103 & 8.711 & 5.299\\
Regression (Pairwise Interactions) & 8.452 & 5.331 & 8.977 & 4.616 & 7.578 & 4.070 & 8.600 & 4.921\\
Regression (AIC) & 7.936 & 4.813 & 9.008 & 4.634 & 7.356 & 4.028 & 8.530 & 4.587\\
Regression (BIC) & 7.778 & 4.707 & 9.632 & 4.685 & 7.263 & 4.038 & 9.099 & 4.556\\
Bayesian Mixture & 7.379 & 4.315 & 8.182 & 4.002 & 7.177 & 3.572 & 7.777 & 4.049\\
\bottomrule
\end{tabular}
\begin{tablenotes}
\small
\item \textit{Note.} Results are based on 200 replicates per simulation setting, for the missed fatality total $D_0$ and the missed incident count $N_0$. Relative error is calculated as $(\hat x -x_{\mathrm{true}})/x_{\mathrm{true}}$. For the Bayesian mixture model, point estimates correspond to posterior medians. `Overall' refers to results computed across all three scenarios jointly.
\end{tablenotes}
\end{threeparttable}
\end{table}

\section{Prior Sensitivity Analysis}
\label{sec:prior-sensitivity}

Given the known sensitivity of multiple-systems estimation to parameter choices, model selection, and prior assumptions, we conduct two sets of robustness checks. First, we report a simple regression baseline alongside the mixture posteriors across all list configurations and stratifications (see Fig.~\ref{fig:main_results}). Second, to investigate the influence of prior choices, we estimate our baseline stratified model (four lists, year $\times$ season strata) across a total of 243 different prior specifications, all of which reflect reasonable \textit{a priori} assumptions within a neighborhood of our baseline setup. Specifically, we vary five factors on a full $3^5$ grid:
(i) the tail weight of the Beta prior on $p^{\mathrm{nb}}_k$, with $b_\omega \in \{2.5, 3, 5\}$ ranging from a heavier to a lighter upper tail of the implied odds $\omega_k$, and $a_\omega$ rescaled in each case according to the mean-preserving centering rule of Eq.~(\ref{eq:centring}) so that the prior predictive mean mark is held fixed at $2G - 1$ across the grid;
(ii) the prior on the negative binomial size, $r_k \sim \mathrm{Gamma}(2,1)$ (default), $\mathrm{Gamma}(1,1)$ (diffuse, favoring strongly overdispersed components), and $\mathrm{Gamma}(4,2)$ (tighter around the same mean);
(iii) the concentration prior, $\alpha_g \sim \mathrm{Gamma}(2,2)$, $\mathrm{Gamma}(2,1)$, and $\mathrm{Gamma}(2,0.5)$, with prior means $1$, $2$, and $4$, tracing out a sparse-to-flexible ladder;
(iv) the capture probability prior, $p_{kj} \sim \mathrm{Beta}(1,1)$ (uniform), $\mathrm{Beta}(1,2)$ (weakly favoring low capture), and $\mathrm{Beta}(2,2)$ (mildly centered);
and (v) the number of mixture components, $K \in \{80, 100, 120\}$.

\begin{figure}[!t]
    \centering

    \begin{subfigure}[b]{0.5\textwidth}
        \centering
        \includegraphics[width=\textwidth, trim={0 0 0 0}, clip]{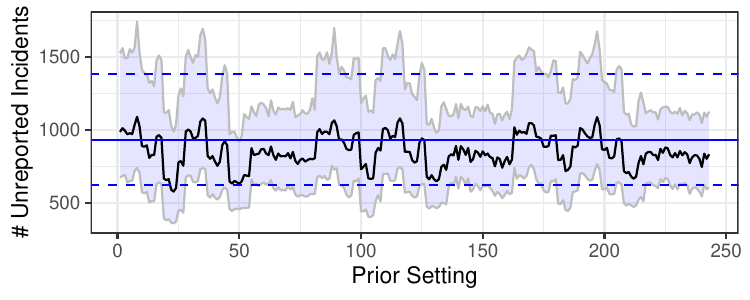}
        \caption{Unreported incidents.}
    \end{subfigure}%
    \begin{subfigure}[b]{0.5\textwidth}
        \centering
        \includegraphics[width=\textwidth, trim={0 0 0 0}, clip]{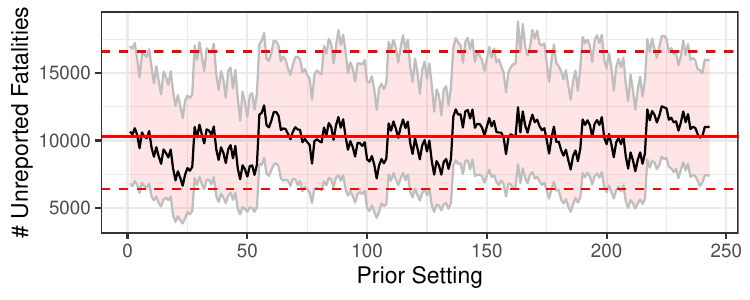}
        \caption{Unreported fatalities.}
    \end{subfigure}

    \caption{Prior sensitivity analysis for the baseline four-list, year $\times$ season model. For each of 243 alternative prior specifications, estimated posterior median, 2.5\% quantile and 97.5\% quantile are shown. Horizontal lines indicate estimates under the baseline prior specification.}
    \label{fig:prior_sensitivity}
\end{figure}

The results are provided in Fig.~\ref{fig:prior_sensitivity}, where the estimated posterior quantiles of missed incidents and fatalities for all configurations are shown and compared to the baseline specification. We find that while there is naturally some variation in the estimates, the overall deviations remain relatively small. Moreover, the 95\% credible interval of the baseline specification provides a reasonable summary of the overall variation across prior configurations. The posterior medians of the alternative specifications largely fall within this interval, so the reported baseline uncertainty is broadly representative of prior uncertainty as well.

\section{Exploratory Analysis of Migrant Mortality Rates}
\label{sec:mort_est}

We aim to provide an estimate of mortality \textit{rates} for migrants using the Central Mediterranean route, complementing the fatality \textit{counts} focused on in the main manuscript. Deriving these rates is challenging, as they are typically defined by deaths over attempted crossings. Attempted crossings are the sum of (i) successful arrivals, (ii) fatalities en route, and (iii) interceptions by North African coast guards. While (i) is measured relatively well by reception systems in Europe, components (ii) and (iii) are largely unobserved, making the denominator -- and thus the mortality rate -- highly uncertain. Formally, the mortality rate $\text{MR}$ is defined as
\begin{equation}
\label{eq:mortrate}
    \text{MR} = \frac{F}{F+A+I} = (1-\rho)\frac{F}{A+F}
\end{equation}
where $F$ denotes fatalities, $A$ denotes arrivals, $I$ denotes interceptions and $\rho$ represents the rate of interception (i.e., the proportion of total attempts that result in interception). We aim to recover a probability distribution for $\text{MR}$ based on this relationship. For the number of fatalities $F$, we rely on the posterior distribution derived from our baseline model specification, aggregating the year $\times$ season draws to annual totals within each retained MCMC draw so that the cross-year posterior dependence is preserved. For arrivals $A$, we use the number of registered border crossings via the Central Mediterranean route as reported by \textit{Frontex}.\footnote{Available from \url{https://www.frontex.europa.eu/what-we-do/monitoring-and-risk-analysis/migratory-map/} (accessed Feb. 9, 2026). Note that, as with the fatality record, coverage of 2014 is incomplete (the MMP series begins on February 17, 2014), so 2014 figures refer to a partial year.}

Modeling the coast guard interception rate, $\rho$, requires assumptions. This is not straightforward, as reliable data on interceptions are virtually nonexistent. We therefore adopt a two-scenario approach. In a first scenario (`uniform'), we let $\rho$ follow a uniform distribution on $[0, 1]$, constant across all years, which we consider as a broad reference distribution. In the second scenario (`calibrated'), we calibrate a distribution for $\rho$ based on additional data collected by the IOM.\footnote{Available from \url{https://data.humdata.org/dataset/iom-missing-migrants-project-data} (accessed Feb. 19, 2026).} These suggest that monthly interception rates between 2016 and 2024 -- after removing outliers -- vary approximately between 3\% and 67\%. However, these data are highly uncertain, and typically based on lower bound fatality figures. In addition, reports from North African coast guards are irregular, rarely transparent, and may suffer from reporting biases. 
Consequently, we assume a relatively conservative $\text{Beta}(1.3, 2.8)$ distribution for $\rho$. This specification has a mean interception share of approximately 31.7\% and a median of approximately 28.5\%, with roughly 95\% of the probability mass falling between 2\% and 77\%, providing a broad coverage of plausible values. We assume this distribution is constant across years, drawing one common interception share per simulated 2014--2025 trajectory.\footnote{Interception dynamics likely shifted over time--particularly following a memorandum of understanding between Italy and Libya in 2017. Further exploratory exercises based on separate dynamics pre- and post-2017 did, however, yield comparable results. We leave a more granular temporal investigation for future work.}

\begin{figure}[t]
\footnotesize
    \centering
    \begin{subfigure}[b]{0.5\textwidth}
        \centering
        \includegraphics[width=\textwidth, trim={0 0 0 0}, clip]{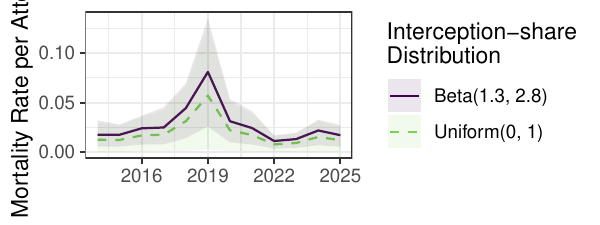}
        \caption{Mortality rates (by year).}
    \end{subfigure}%
    \begin{subfigure}[b]{0.5\textwidth}
        \centering
        \includegraphics[width=\textwidth, trim={0 0 0 0}, clip]{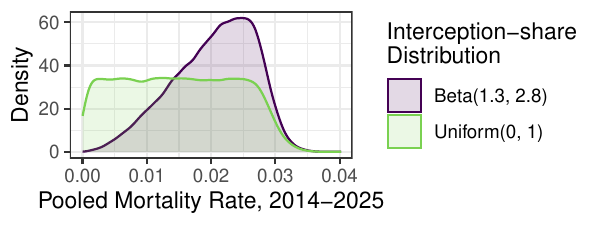}
        \caption{Pooled period mortality rates.}
    \end{subfigure}\\
    \begin{subfigure}{0.4\textwidth}
    \centering
        \includegraphics[width=\linewidth]{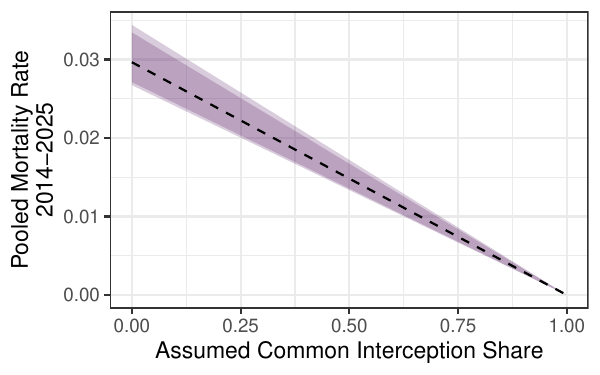}
    \caption{Mortality rates (varying $\rho$ in $[0,1]$).}
    \end{subfigure}
    \caption{Monte Carlo-based estimates of mortality rates. For top panels, interception probability is either assumed uniform or arising from a calibrated Beta distribution; see text for details. Ribbons in panel (a) are pointwise 95\% simulation intervals per year. Bottom panel shows estimates based on a grid of fixed common values for $\rho$. Point estimates refer to the posterior median. Shaded areas are 2.5\% and 97.5\% quantiles.}
    \label{fig:mortality_rates}
\end{figure}

Based on these assumptions, we simulate mortality rates via Monte Carlo methods. Each of the 50,000 retained joint posterior trajectories for annual fatalities is paired with a draw of $\rho$, substituted into equation (\ref{eq:mortrate}), and the resulting samples for $\text{MR}$ are collected. As the period-level summary we report the pooled mortality rate, i.e., total fatalities over total attempted crossings across 2014--2025. The top panels of Fig.~\ref{fig:mortality_rates} summarize the resulting distributions, both stratified by year and pooled over the period. Overall, estimates range up to mortality rates of $3.5\%$ in the upper tail. Estimates remain relatively stable over time, with the exception of 2019, where significantly lower arrival numbers imply a higher mortality rate. However, the wide uncertainty bounds prevent definitive conclusions regarding this spike. As a robustness check, the bottom panel of Fig.~\ref{fig:mortality_rates} reports the implied mortality rate distributions for a grid of fixed common values of $\rho$. Again, under the lowest-interception rate scenarios, mortality rates rise to around 3.5\%.

In general, this analysis should be interpreted with caution given the reliance on unverifiable assumptions and model-based estimates. Nevertheless, it is worth noting that our posterior densities broadly align with simpler, alternative approaches. Related literature typically estimates rates of similar magnitude, ranging from 0.3\% to 1.8\%\footnote{See \url{https://www.bertelsmann-stiftung.de/fileadmin/files/Projekte/Migration_fair_gestalten/IB_PolicyBrief_2018_06_Asylum_Centers.pdf} (accessed Feb. 9, 2026).} or, in direct IOM estimates,\footnote{ \url{https://publications.iom.int/system/files/pdf/mortality-rates.pdf} (accessed Feb. 9, 2026).} averaging around 2.8\% with a similar observed increase in 2019. \citet{steinhilper2018contested} report a point estimate of 0.875\% for \textit{all} Mediterranean irregular migration routes for the period 2009-2016, ignoring coast guard interceptions. Note that the analysis presented here applies only to the Central Mediterranean Route and only for the years 2014--2025. It should not be interpreted as applying to other routes or time periods.

\section{Additional Figures and Tables}
\subsection*{}

\begin{figure}[!h]
    \centering
    \begin{subfigure}[b]{0.5\textwidth}
        \centering
        \includegraphics[width=\textwidth, trim={0 0 0 0}, clip]{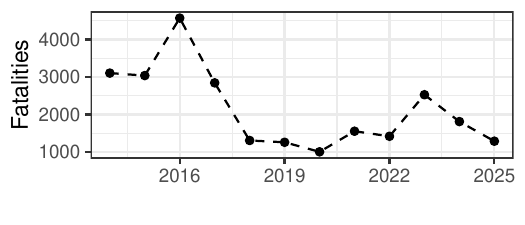}
        \caption{Total fatalities.}
    \end{subfigure}\hfill
    \begin{subfigure}[b]{0.5\textwidth}
        \centering
        \includegraphics[width=\textwidth, trim={0 0 0 0}, clip]{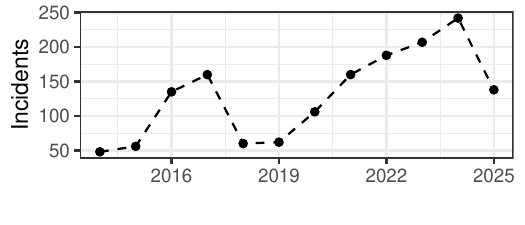}
        \caption{Total incidents.}
    \end{subfigure}\\
    \begin{subfigure}[b]{0.5\textwidth}
        \centering
        \includegraphics[width=\textwidth, trim={0 0 0 0}, clip]{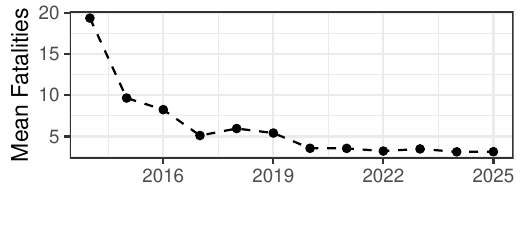}
        \caption{Mean fatalities per incident.}
    \end{subfigure}\hfill
        \begin{subfigure}[b]{0.5\textwidth}
        \centering
        \includegraphics[width=\textwidth, trim={0 0 0 0}, clip]{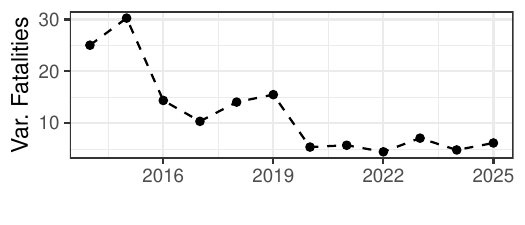}
        \caption{Variance of fatalities per incident.}
    \end{subfigure}\\
       \begin{subfigure}[b]{\textwidth}
        \centering
        \includegraphics[width=0.5\textwidth, trim={0 0 0 0}, clip]{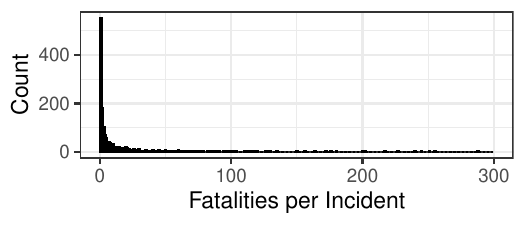}
        \caption{Fatalities per incident.}
    \end{subfigure}

    \caption{Panels (a) and (b) show total fatalities and incidents over time. Panels (c) and (d) track the sample mean and variance of fatalities over time. Panel (e) shows a histogram of fatalities per incident, omitting eight high fatality incidents for visualization purposes.}
    \label{fig:incidents_fatalities_over_time}
\end{figure}

\begin{figure}[t]
    \centering
    \begin{subfigure}[b]{0.75\textwidth}
        \centering
        \includegraphics[width=\textwidth, trim={0 0 0 0}, clip]{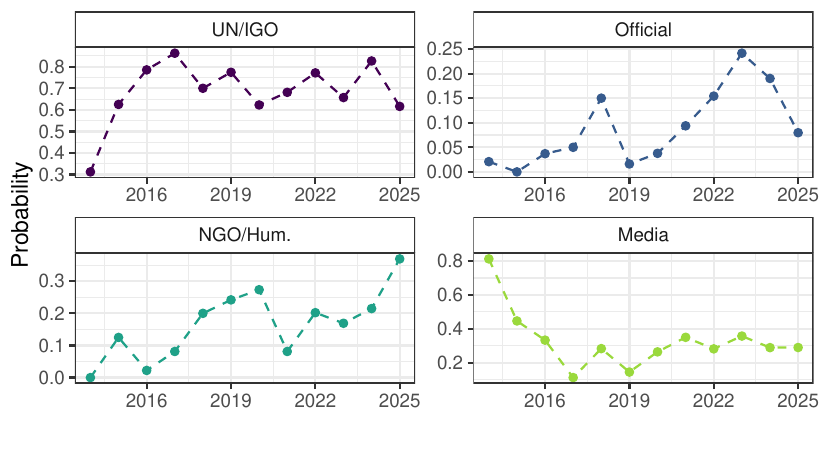}
        \caption{Marginal reporting by list.}
    \end{subfigure}\\
    \begin{subfigure}[b]{0.8\textwidth}
        \centering
        \includegraphics[width=\textwidth, trim={0 0 0 0}, clip]{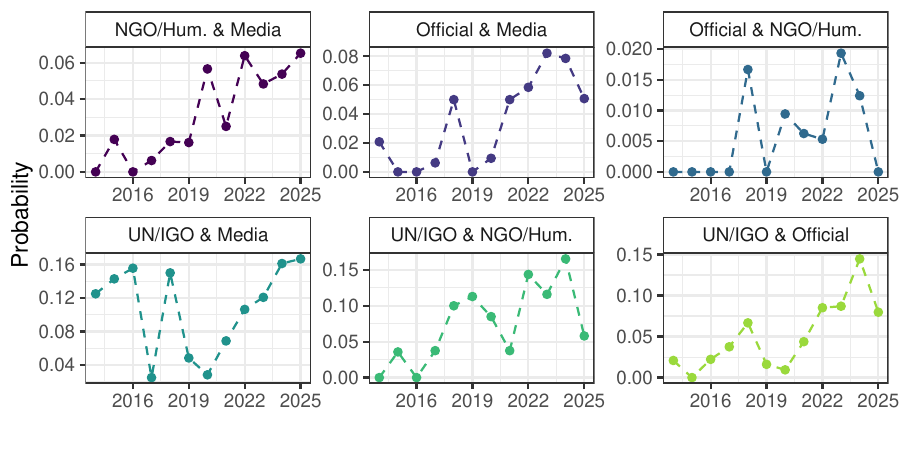}
        \caption{Pairwise reporting by list pair.}
    \end{subfigure}\\
    \begin{subfigure}[b]{0.75\textwidth}
        \centering
        \includegraphics[width=\textwidth, trim={0 0 0 0}, clip]{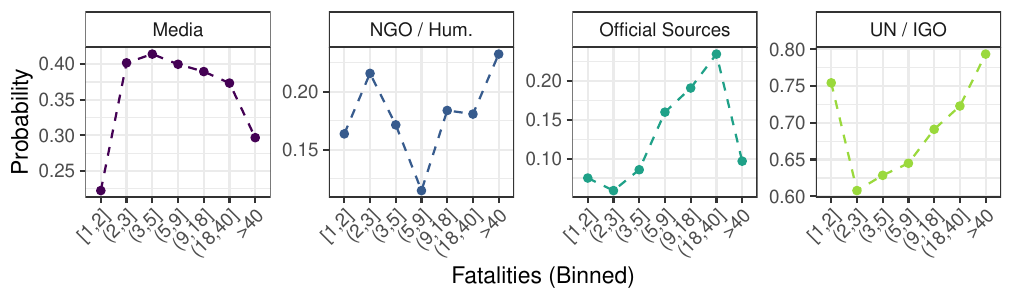}
        \caption{Reporting by incident severity.}
        \label{fig:probability_by_fatality}
    \end{subfigure}

    \caption{Panel (a) shows marginal reporting patterns by list. Panel (b) shows pairwise reporting patterns for list pairs. Panel (c) shows reporting patterns as a function of fatalities. The $x$-axis displays fatality bins based on deciles of the empirical fatality distribution. Figures are based on observed data sample and are conditional on an incident being observed.}
    \label{fig:reporting_by_list}
\end{figure}

\begin{figure}[htbp]
    \centering

    \begin{subfigure}{0.5\textwidth}
        \centering
        \includegraphics[width=\textwidth, trim={0 0 0 0}, clip]{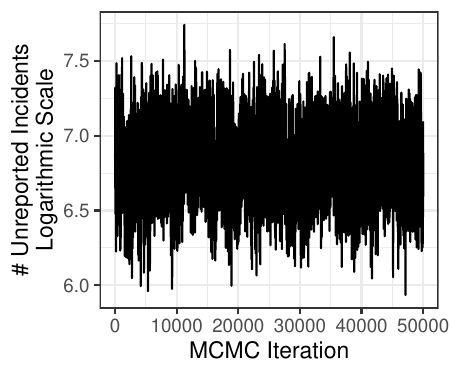}
        \caption{Log Missing Incidents.}
    \end{subfigure}%
    \begin{subfigure}{0.5\textwidth}
        \centering
        \includegraphics[width=\textwidth, trim={0 0 0 0}, clip]{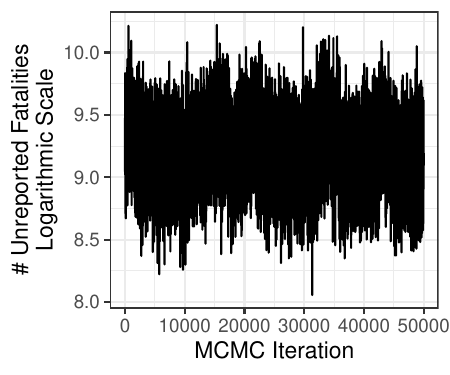}
        \caption{Log Missing Fatalities.}
    \end{subfigure}

    \caption{MCMC trace plots of total unreported incidents and fatalities under the baseline four-list, year $\times$ season specification (totals aggregated across strata within each retained draw).}
    \label{fig:trace}
\end{figure}


\begin{figure}
    \centering
    \includegraphics[width=\linewidth]{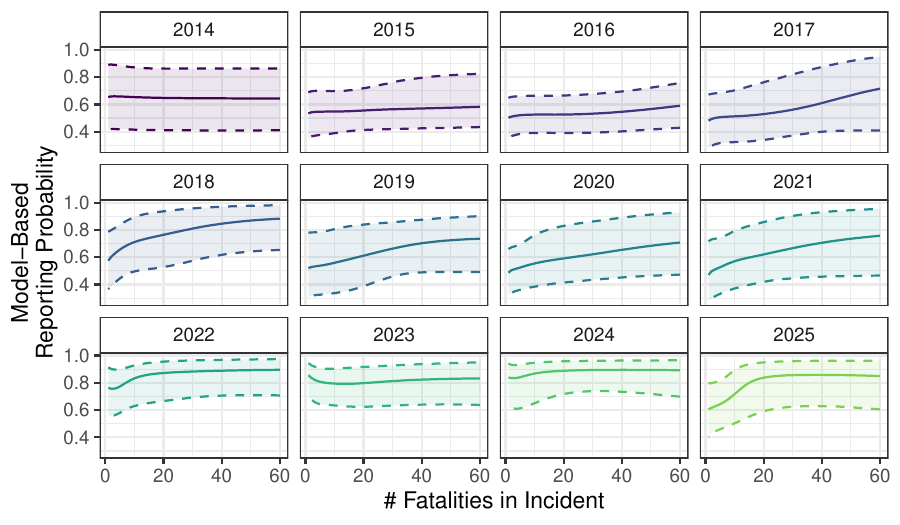}
    \caption{Model-based probability of an incident being reported ($y$-axis) as a function of fatalities in incident ($x$-axis) over years (panels). Annual curves are obtained by mixing the two seasonal strata of each year within every draw, weighted by each stratum's share of that year's total incidents (observed plus imputed), approximating the detection probability of a randomly selected incident of the given severity in that year. Dashed lines are posterior median estimates, solid lines are 95\% credible intervals.}
    \label{fig:fatality_reporting_time}
\end{figure}

\begin{figure}
    \centering
    \includegraphics[width=\linewidth]{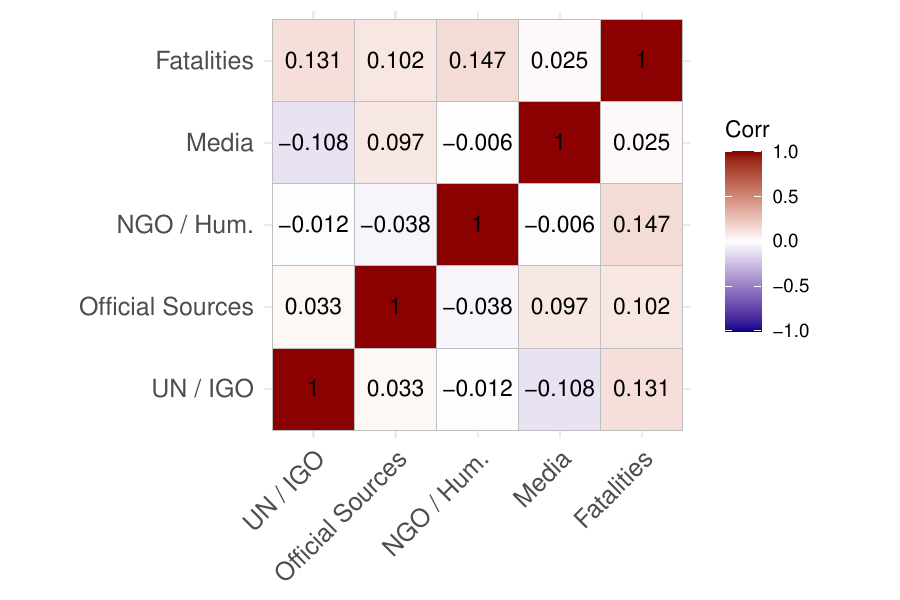}
    \caption{Posterior mean estimate of raw correlation matrix of reporting patterns $\mathbf{s}_i$ and fatalities $y_i$, based on complete data (including imputed values for unobserved incidents). Results are averaged across year $\times$ season strata.}
    \label{fig:correlation}
\end{figure}
\clearpage
\vspace{5cm}
\begin{table}
\centering
\footnotesize
\setlength{\tabcolsep}{15pt} 
\renewcommand{\arraystretch}{0.6}
\begin{threeparttable}
\caption{List Overlap Patterns (Four Lists).}
\label{tab:mse_overlap}
\begin{tabular}{ccccr}
\toprule
UN/IGO & Official & NGO/Hum. & Media & Count \\
\midrule
$\bullet$ &  &  &  & 771\\
 & $\bullet$ &  &  & 35\\
 &  & $\bullet$ &  & 97\\
 &  &  & $\bullet$ & 236\\
$\bullet$ & $\bullet$ &  &  & 72\\
$\bullet$ &  & $\bullet$ &  & 109\\
$\bullet$ &  &  & $\bullet$ & 119\\
 & $\bullet$ & $\bullet$ &  & 3\\
 & $\bullet$ &  & $\bullet$ & 33\\
 &  & $\bullet$ & $\bullet$ & 28\\
$\bullet$ & $\bullet$ & $\bullet$ &  & 1\\
$\bullet$ & $\bullet$ &  & $\bullet$ & 28\\
$\bullet$ &  & $\bullet$ & $\bullet$ & 23\\
 & $\bullet$ & $\bullet$ & $\bullet$ & 5\\
$\bullet$ & $\bullet$ & $\bullet$ & $\bullet$ & 2\\
\midrule
\multicolumn{4}{l}{\textit{Total}} & 1,562 \\
\bottomrule
\end{tabular}
\begin{tablenotes}
\small
\item \textit{Note:} Each row represents a unique capture pattern across the four data sources. `$\bullet$' indicates inclusion in the respective list. Count refers to the number of records observed in each overlap pattern. UN/IGO = United Nations/Intergovernmental Organizations; NGO/Hum.\ = Non-Governmental Organizations/Humanitarian sources.
\end{tablenotes}
\end{threeparttable}
\end{table}

\begin{table}[t]
\centering
\footnotesize
\setlength{\tabcolsep}{15pt} 
\renewcommand{\arraystretch}{0.6}
\begin{threeparttable}
\caption{List Overlap Patterns (Three Lists).}
\label{tab:mse_overlap_3L}
\begin{tabular}{cccr}
\toprule
UN/IGO/NGO/Hum. & Official & Media & Count \\
\midrule
$\bullet$ &  &  & 977\\
 & $\bullet$ &  & 35\\
 &  & $\bullet$ & 236\\
$\bullet$ & $\bullet$ &  & 76\\
$\bullet$ &  & $\bullet$ & 170\\
 & $\bullet$ & $\bullet$ & 33\\
$\bullet$ & $\bullet$ & $\bullet$ & 35\\
\midrule
\multicolumn{3}{l}{\textit{Total}} & 1,562 \\
\bottomrule
\end{tabular}
\begin{tablenotes}
\small
\item \textit{Note:} Each row represents a unique capture pattern across the three data sources. `$\bullet$' indicates inclusion in the respective list. Count refers to the number of records observed in each overlap pattern. UN/IGO/NGO/Hum = United Nations/Intergovernmental Organizations/Non-Governmental Organizations/Humanitarian sources.
\end{tablenotes}
\end{threeparttable}
\end{table}

\begin{table}[!htbp]
\centering
\footnotesize
\setlength{\tabcolsep}{9pt} 
\renewcommand{\arraystretch}{0.6}
\begin{threeparttable}
\caption{List Overlap Patterns (Five Lists).}
\label{tab:mse_overlap_5L}
\begin{tabular}{cccccr}
\toprule
UN/IGO & Official EU & Official Other & NGO/Hum. & Media & Count \\
\midrule
$\bullet$ &  &  &  &  & 771\\
 &  &  &  & $\bullet$ & 236\\
 &  &  & $\bullet$ &  & 97\\
 &  & $\bullet$ &  &  & 30\\
 & $\bullet$ &  &  &  & 5\\
$\bullet$ &  &  &  & $\bullet$ & 119\\
$\bullet$ &  &  & $\bullet$ &  & 109\\
$\bullet$ &  & $\bullet$ &  &  & 63\\
 &  & $\bullet$ &  & $\bullet$ & 32\\
 &  &  & $\bullet$ & $\bullet$ & 28\\
$\bullet$ & $\bullet$ &  &  &  & 9\\
 & $\bullet$ &  & $\bullet$ &  & 2\\
 & $\bullet$ &  &  & $\bullet$ & 1\\
 &  & $\bullet$ & $\bullet$ &  & 1\\
$\bullet$ &  & $\bullet$ &  & $\bullet$ & 25\\
$\bullet$ &  &  & $\bullet$ & $\bullet$ & 23\\
 &  & $\bullet$ & $\bullet$ & $\bullet$ & 4\\
$\bullet$ & $\bullet$ &  &  & $\bullet$ & 3\\
$\bullet$ &  & $\bullet$ & $\bullet$ &  & 1\\
 & $\bullet$ &  & $\bullet$ & $\bullet$ & 1\\
$\bullet$ & $\bullet$ & $\bullet$ & $\bullet$ & $\bullet$ & 2\\
\midrule
\multicolumn{5}{l}{\textit{Total}} & 1,562 \\
\bottomrule
\end{tabular}
\begin{tablenotes}
\small
\item \textit{Note:} Each row represents a unique capture pattern across the five data sources, where official sources are split into European (Official EU) and non-European (Official Other) institutions. `$\bullet$' indicates inclusion in the respective list. Count refers to the number of records observed in each overlap pattern. UN/IGO = United Nations/Intergovernmental Organizations; NGO/Hum.\ = Non-Governmental Organizations/Humanitarian sources.
\end{tablenotes}
\end{threeparttable}
\end{table}

\end{document}